\documentclass[pdflatex,sn-mathphys-num]{sn-jnl}

\usepackage{graphicx}%
\usepackage{multirow}%
\usepackage{amsmath,amssymb,amsfonts}%
\usepackage{amsthm}%
\usepackage{mathrsfs}%
\usepackage[title]{appendix}%
\usepackage{xcolor}%
\usepackage{textcomp}%
\usepackage{manyfoot}%
\usepackage{booktabs}%
\usepackage{algorithm}%
\usepackage{algorithmicx}%
\usepackage{algpseudocode}%
\usepackage{listings}%
\usepackage{csquotes}
\usepackage[T1]{fontenc}
\usepackage{epsfig}
\usepackage{amssymb}
\usepackage{graphicx}
\usepackage{subcaption}
\usepackage{multirow}
\usepackage{tabularx}
\usepackage[justification=centering]{caption}
\usepackage{amsthm}
\usepackage{multicol, blindtext}
\usepackage{float}
\usepackage{lineno}
\usepackage{enumitem}
\usepackage{pdfpages}
\usepackage{booktabs}
\usepackage{csquotes}
\usepackage{longtable}
\usepackage{adjustbox}
\usepackage{lineno}
\usepackage{float}

\theoremstyle{thmstyleone}%
\theoremstyle{thmstyletwo}%

\theoremstyle{thmstylethree}%

\begin{document}
\title[Artificial Intelligence in Rural Healthcare Delivery]{Artificial Intelligence in Rural Healthcare Delivery: Bridging Gaps and Enhancing Equity through Innovation}

\author*[1]{\fnm{Kiruthika} \sur{Balakrishnan}}\email{bkiruthi@uic.edu}

\author[2]{\fnm{Durgadevi} \sur{Velusamy}}

\author[1]{\fnm{Hana E.} \sur{Hinkle}}

\author[1]{\fnm{Zhi} \sur{Li}}

\author[3]{\fnm{Karthikeyan} \sur{Ramasamy}}

\author[4]{\fnm{Hikmat} \sur{Khan}}

\author[5]{\fnm{Srini} \sur{Ramaswamy}}

\author[6,7]{\fnm{Pir Masoom} \sur{Shah}}

\affil*[1]{\orgname{University of Illinois College of Medicine Rockford}, \orgdiv{National Center for Rural Health Professions}, \orgaddress{\street{1601 Parkview Avenue}, \city{Rockford}, \postcode{61107}, \state{IL}, \country{USA}}}

\affil[2]{\orgname{Sri Sivasubramaniya Nadar College of Engineering}, \orgdiv{Department of Information Technology}, \orgaddress{\city{Kalavakkam}, \postcode{603110}, \country{India}}}

\affil[3]{\orgname{M. Kumarasamy College of Engineering}, \orgdiv{Department of Electrical and Electronics Engineering}, \orgaddress{\city{Karur}, \postcode{639113}, \country{India}}}

\affil[4]{\orgname{The Ohio State University Wexner Medical Center}, \orgdiv{College of Medicine, Department of Pathology}, \orgaddress{\city{Columbus}, \state{OH}, \postcode{43210}, \country{USA}}}

\affil[5]{\orgname{iWorks Corporation}, \orgaddress{\city{Reston}, \state{VA}, \postcode{20191}, \country{USA}}}

\affil[6]{\orgname{Central South University}, \orgdiv{School of Computer Science and Engineering}, \orgaddress{\city{Changsha}, \postcode{410010}, \country{China}}}

\affil[7]{\orgname{Bacha Khan University}, \orgdiv{Department of Computer Science}, \orgaddress{\city{Charsadda}, \country{Pakistan}}}

\abstract{Rural healthcare faces persistent challenges, including inadequate infrastructure, workforce shortages, and socioeconomic disparities that hinder access to essential services. This study investigates the transformative potential of artificial intelligence (AI) in addressing these issues in underserved rural areas. We systematically reviewed 109 studies published between 2019 and 2024 from PubMed, Embase, Web of Science, IEEE Xplore, and Scopus. Articles were screened using PRISMA guidelines and Covidence software. A thematic analysis was conducted to identify key patterns and insights regarding AI implementation in rural healthcare delivery. The findings reveal significant promise for AI applications, such as predictive analytics, telemedicine platforms, and automated diagnostic tools, in improving healthcare accessibility, quality, and efficiency. Among these, advanced AI systems, including Multimodal Foundation Models (MFMs) and Large Language Models (LLMs), offer particularly transformative potential. MFMs integrate diverse data sources, such as imaging, clinical records, and bio signals, to support comprehensive decision-making, while LLMs facilitate clinical documentation, patient triage, translation, and virtual assistance. Together, these technologies can revolutionize rural healthcare by augmenting human capacity, reducing diagnostic delays, and democratizing access to expertise. However, barriers remain, including infrastructural limitations, data quality concerns, and ethical considerations. Addressing these challenges requires interdisciplinary collaboration, investment in digital infrastructure, and the development of regulatory frameworks. This review offers actionable recommendations and highlights areas for future research to ensure equitable and sustainable integration of AI in rural healthcare systems.}

\maketitle

\section{Introduction}
Globally, rural populations face substantial health inequities, with nearly 24\% (2 billion) people lacking adequate access to primary health services. This is often due to systemic challenges, such as limited healthcare infrastructure in certain regions, insufficient workforce capacity, geographic isolation, and socioeconomic disparities \cite{WHO}. These health inequities are characterized by a severe shortage and unequal distribution of health workers, with only 36\% of the global nursing workforce serving rural areas despite housing nearly half of the world's population \cite{WHO}. These inequities may be due to persistent challenges in rural healthcare, including low population density, limited economies of scale, and difficulties in recruiting and retaining healthcare professionals \cite{Andrew}. Additionally, uneven infrastructure development and centralization of services force rural residents to travel long distances for care, exacerbating access disparities \cite{Andrew}. Digital connectivity gaps further hinder the implementation of telehealth solutions, leaving rural communities particularly vulnerable during crises like the COVID-19 pandemic and extreme weather events \cite{Andrew}. Addressing these challenges is critical for achieving rural global health equity and ensuring the well-being of nearly 41.5\% of the world's population who reside in rural or remote areas. To address these challenges, innovative approaches such as artificial intelligence (AI) offer promising avenues for enhancing accessibility, resilience, and continuity of care in rural healthcare systems \cite{mapari}.

Rapid advancements in artificial intelligence (AI) offer innovative opportunities to address healthcare challenges and reduce health inequities, particularly in rural settings \cite{mapari}. AI, a dynamic field of computer science, simulates human intelligence processes such as learning, reasoning, and decision-making to achieve more efficient outcomes \cite{Krittanawong}. It has facilitated the development of cutting-edge tools and techniques, including machine learning (ML) \cite{Andishgar}, expert systems \cite{Ahmed}, speech recognition \cite{Oluw}, natural language processing (NLP) \cite{Abu1}, predictive analytics \cite{Abu}, and computer vision \cite{Jones}. These tools have been successfully applied across various domains, such as communication security \cite{DV}, optimization\cite{kb,rk}, predictive modeling\cite{arivoli,choka}, smart grids\cite{DV1}, and automatic disease diagnosis\cite{ura}.

In healthcare, numerous studies have demonstrated the effectiveness of ML algorithms in diagnosing and classifying conditions, such as coronary artery disease\cite{DV2}, obstructive sleep apnea\cite{surrel}, cancer\cite{hosni}, Alzheimer's disease\cite{dpan}, liver disease \cite{kb1} and chronic kidney disease\cite{song}. These advancements hold significant potential to transform rural healthcare delivery. By addressing critical gaps in infrastructure, workforce shortages, and access to essential services, AI-driven solutions can enhance diagnostic accuracy, streamline treatment planning, improve health outcomes, and reduce costs \cite{Alowais}. Together, these innovations offer a transformative approach to overcoming persistent challenges in rural healthcare systems.

Building on these advancements, AI in healthcare has demonstrated the capacity to manage vast amounts of multimedia data, including electronic medical records (EMR) and personal health records (PHR), enabling precise analysis and decision-making \cite{Dash}. AI systems improve operational efficiency, reduce medical errors, and empower healthcare providers by assisting clinical workflows, particularly in resource-constrained rural settings \cite{Guo}. Furthermore, during crises like the COVID-19 pandemic, AI effectively supported diagnostics, optimized resource allocation, and alleviated physician workloads \cite{Wu}. These capabilities are significant in rural healthcare settings, where crises often exacerbate resource constraints and workforce shortages. Despite these promising developments, there remains a notable gap in studies focused on tailoring AI applications to rural healthcare delivery, where unique challenges like limited infrastructure, ethical considerations, a lack of legal frameworks, and data availability and security issues persist. While many studies focus on telehealth and general AI advancements, these often lack the specificity for low-resource and rural settings. Some articles explore AI applications in conventional healthcare environments or urban areas, but their relevance to rural healthcare challenges remains uncertain\cite{strik}.

This literature review explores the existing applications, challenges, and gaps in the use of AI in rural healthcare delivery. We examine current research on AI-driven solutions in rural healthcare, identifying how these technologies align with rural healthcare systems' unique needs, preferences, and barriers. Additionally, we discuss the opportunities and limitations of implementing AI in these settings, highlighting its potential to improve rural healthcare by addressing persistent inequities and challenges. This review also delves into the current state of AI applications in healthcare, the datasets and methodologies employed, and the key considerations for advancing AI-driven innovations tailored to rural contexts. Finally, we outline critical areas for future research to ensure that AI solutions effectively address the diverse challenges of rural healthcare delivery.

\section{Methods}
This literature review followed the Cochrane guidelines for systematic reviews to ensure a thorough, transparent, and unbiased methodology. Additionally, the Preferred Reporting Items for Systematic Reviews and Meta-Analyses for Scoping Reviews (PRISMA-ScR) framework was applied to maintain structure and consistency in reporting. Adherence to these internationally recognized standards enhances the review's credibility by ensuring a systematic process for identifying, selecting, and synthesizing relevant studies. This study relies on existing literature and does not involve data collection from human participants, so it did not require Institutional Review Board (IRB) approval. The research analyzes publicly available, peer-reviewed studies in compliance with ethical standards for non-human subject research. The review methodology establishes the research question and outlines the strategy, including information sources, search queries, quality evaluation based on eligibility criteria, study selection, data extraction, and data synthesis.

\subsection{Information Sources}
This study comprehensively searched peer-reviewed studies across five reputable electronic databases: PubMed, Embase, Web of Science, Scopus, and IEEE Xplore. These databases were chosen for their specialization in various areas of scientific literature, ensuring a broad yet targeted exploration of studies on AI in rural healthcare delivery. The search focused on articles that published between 2019 and 2024 to capture recent advancements and trends in AI applications. We selected this time frame to reflect the latest developments in AI technologies, particularly their growing integration into healthcare delivery systems, which have seen significant progress in recent years. By leveraging the unique strengths of these databases, we ensured a diverse and high-quality selection of studies, enhancing the credibility and robustness of our findings on AI in rural healthcare delivery. This multi-database approach strengthened the validity of our literature review, enabling us to provide valuable insights into the application of AI in rural healthcare delivery.

\subsection{Search strategy}
The search strategy for this review targeted peer-reviewed studies published between 2019 and 2024, focusing on the use of AI in rural healthcare delivery. The search was conducted in November 2024, following PRISMA guidelines. A structured search query utilized a combination of Medical Subject Headings (MeSH) terms, free-text keywords, Boolean operators (\enquote{AND} and \enquote{OR}), and proximity operators to ensure comprehensive and relevant results. The search terms included \enquote{Artificial Intelligence}, \enquote{Machine Learning}, \enquote{Deep Learning}, \enquote{Expert Systems}, \enquote{Fuzzy Logic}, \enquote{Natural Language Processing}, \enquote{Robotics}, and \enquote{Predictive Analytics}, combined with rural healthcare-specific terms such as \enquote{Rural Health}, \enquote{Rural Health Services}, \enquote{Rural Population}, \enquote{Rural Nursing}, \enquote{Rural Hospitals}, and \enquote{Telemedicine}. This systematic strategy facilitated the retrieval of high-quality literature, supporting an in-depth exploration of AI applications in rural healthcare delivery.

\subsection{Eligibility criteria} \label{eligibility}
This review included peer-reviewed journal articles published in English between 2019 and 2024, focusing on the application of AI in rural healthcare delivery. Studies were selected based on their exploration of AI technologies such as machine learning, deep learning, predictive analytics, and AI-enhanced telehealth, addressing challenges in rural settings like access to care, diagnostics, chronic disease management, mental health, preventive care, and health education. Research involving rural healthcare professionals, patients, administrators, or policymakers was prioritized, emphasizing healthcare quality, access, equity, and system efficiency outcomes. Particular attention was given to studies identifying barriers and facilitators for AI implementation in rural healthcare. Exclusion criteria eliminated studies lacking a direct focus on rural healthcare, those addressing AI in healthcare without specific rural relevance, or those unrelated to AI applications in healthcare. Duplicate articles, non-English publications, and non-empirical works such as opinion pieces, commentaries, editorials, reviews, conference abstracts, and documents generated by non-humans (e.g., ChatGpt) were also excluded.

\subsection{Selection and Data Collection}
All records obtained from the database were imported into Covidence, a platform designed to support researchers conducting literature reviews. The selection process involved three stages of screening. In the first stage, duplicate articles were identified and removed. Following this, in the second stage, titles, abstracts, and study types were screened based on predefined eligibility criteria, and articles not meeting the criteria were excluded. The remaining articles underwent full-text screening in the final stage, applying the same eligibility criteria. Articles not meeting the criteria were excluded, with reasons for exclusion documented. Two independent reviewers assessed the studies for eligibility and extracted relevant data from each included study. Any disagreements or uncertainties were resolved through discussion between the reviewers.
Data collection included the article title, author(s), publication year, country of origin, AI application, study objective, methodology, AI algorithms used, performance metrics, clinical practice challenges, benefits, and limitations. The gathered information was systematically organized and documented in a Microsoft Excel spreadsheet to maintain transparency and consistency.

\subsection{Data Synthesis}
The data synthesis involved a qualitative approach to organize and present findings aligned with the study's objectives, analyzing individual studies and the overall dataset to identify patterns, similarities, and key themes. Studies with similar results were grouped under unified concepts to enable a cohesive interpretation, and the quality of each study was meticulously assessed to ensure robustness and credibility. The quality assessment criteria included evaluating whether the study's aims, scope, and content were clearly defined, the research methods and variables were reliable, all questions were answered, findings were clearly stated, the research process was well-documented, conclusions aligned with the study's purpose, and limitations were transparently addressed.

\section{Results}
\subsection{Overview}
In this literature review, 1,691 studies published between 2019 and 2024 were initially identified. After removing 342 duplicate studies, 1,117 were excluded following a title and abstract screening. The remaining 232 studies were subjected to a full-text review based on the inclusion and exclusion criteria outlined in Section \ref{eligibility}. This process resulted in 109 unique studies being included in the review.

\begin{figure}[htbp]
    \centering
    \includegraphics[width=\linewidth]{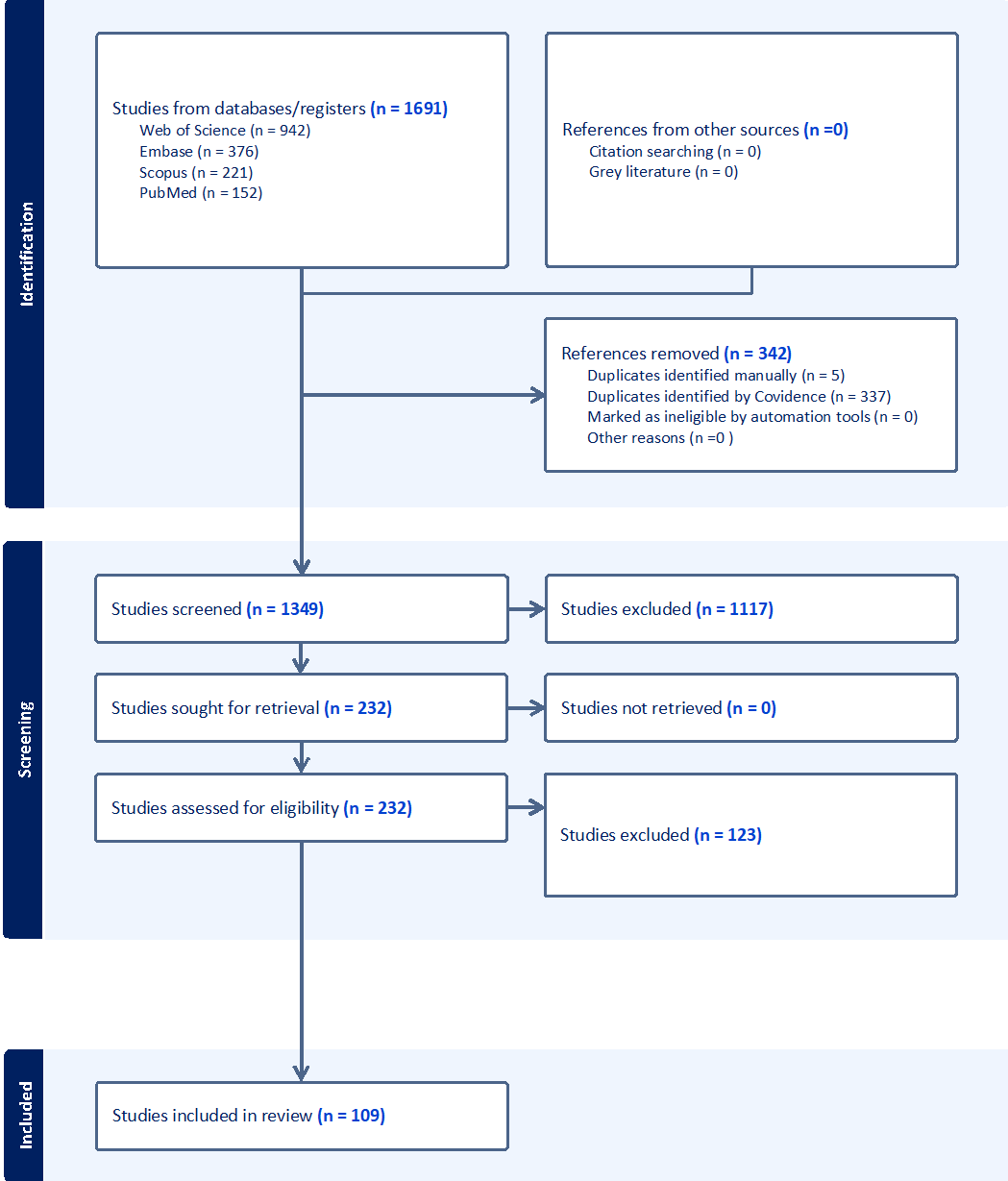}
    \caption{PRISMA flowchart diagram for the proposed review}
    \label{Prisma}
\end{figure}

\begin{figure}[htbp]
    \centering
    \includegraphics[width=\linewidth]{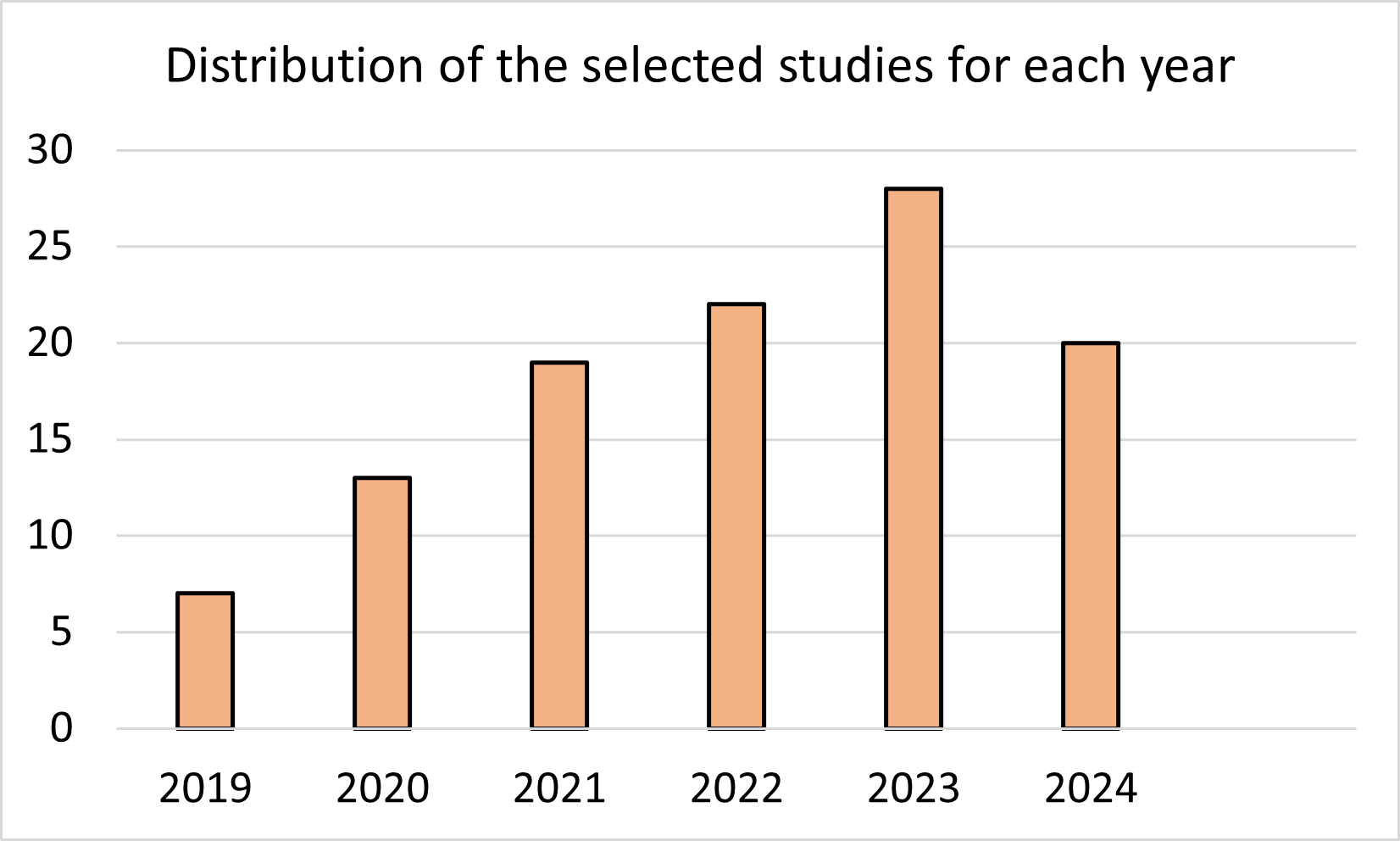}
    \caption{Distribution of the selected studies for each year}
    \label{Distribution}
\end{figure}

Figure \ref{Prisma} presents the PRISMA flowchart, which visualizes the entire screening process. The metadata for the 109 studies included in this review is provided in supplementary section Table 1. The year-wise distribution of the selected studies from various journals is depicted in Figure \ref{Distribution}, illustrating the trends in publications on the use of AI in rural healthcare delivery from 2019 to 2024.

The graph shows 7 papers were published in 2019, 13 in 2020, and 19 in 2021. In 2022, 22 articles were published. The year 2023 recorded the highest number of publications, with 28 papers, while 20 papers were published in 2024.

Geographically, most studies (26\%) were conducted in China, followed by 19\% in India and 15\% in the USA. The most frequently addressed areas of research were diabetic retinopathy care (11\%), cardiac care (9\%), diabetes care (8\%), maternal healthcare (7\%), infectious disease care (6\%), and cancer care (6\%).

A variety of AI techniques were employed across the studies, including deep learning (DL) methods and ensemble techniques such as Random Forest (RF), Gradient Boosting Machines (GBM), Extreme Gradient Boosting (XGB), and AdaBoost. Additionally, ML methods such as Support Vector Machines(SVM), Logistic Regression (LR), and K-Nearest Neighbors (KNN) were widely utilized to enhance diagnostic accuracy, predict chronic disease risks, optimize resource allocation, and support decision-making in rural healthcare delivery.

\subsection{Chronic and Non-Communicable Diseases}

\begin{figure*}
  \centering
  \includegraphics[width=\linewidth]{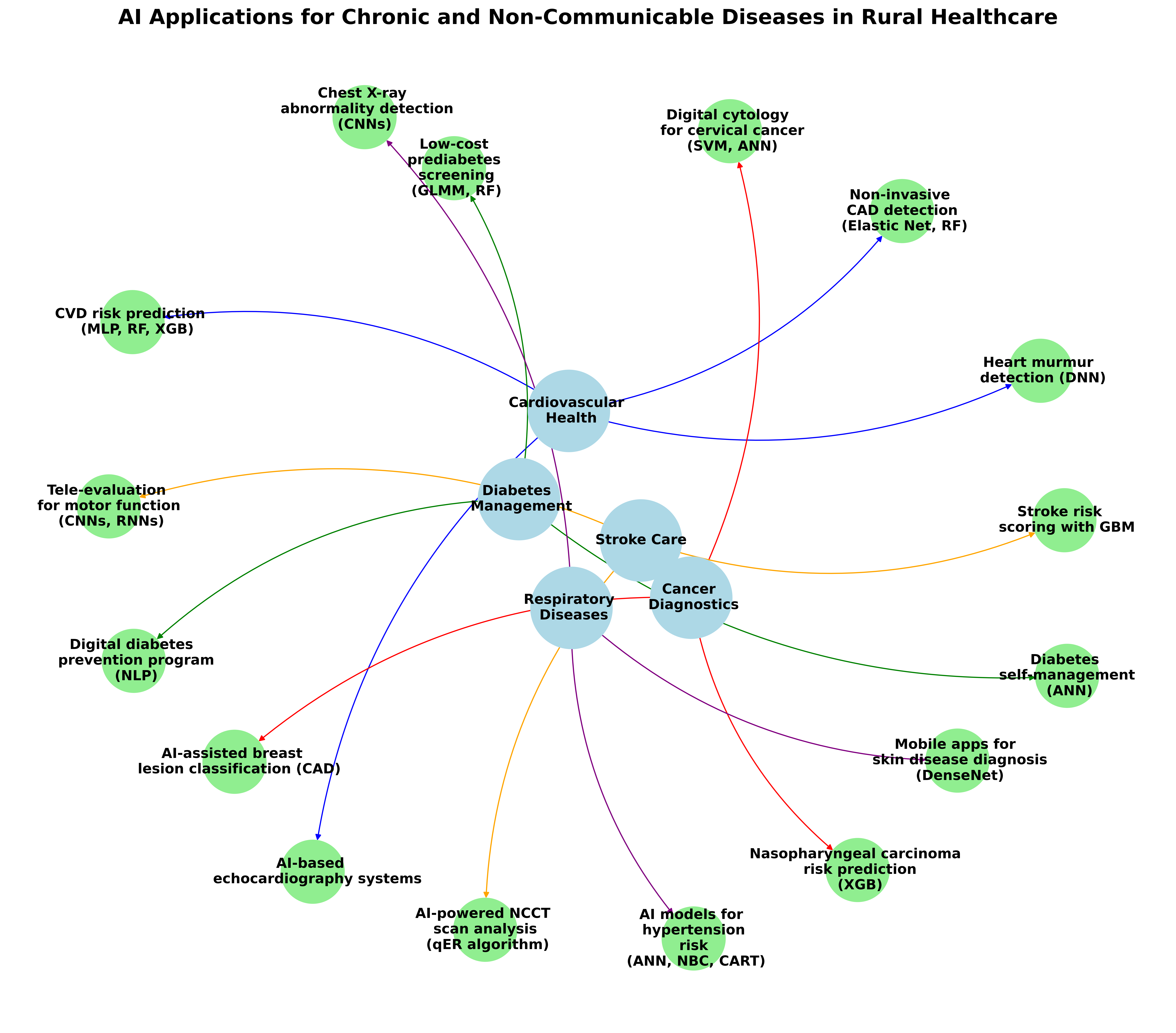}\\
  \caption{AI Applications for Chronic and Non-Communicable Diseases in Rural Healthcare}\label{NCD}
\end{figure*}

Of the 109 studies included in this review, 49 explored AI-based tools for the diagnosis, screening, and management of chronic and non-communicable diseases (NCDs) in rural and underserved populations, as summarized in Figure \ref{NCD}. These studies were categorized into five domains: cardiovascular disease (n = 14), diabetes (n = 11), diabetic retinopathy (DR) (n = 10), cancer (n = 6), and other chronic conditions such as stroke, respiratory, dermatologic, kidney, and women's health (n = 8). Overall, 42 of the 49 studies (86\%) reported quantitative performance metrics such as accuracy, AUC, or F1-score. Accuracy values ranged from 81.2\% to 97.5\%, AUC values from 0.78 to 0.96, and F1-scores from 0.74 to 0.93. RF was the most commonly applied model (n=18), demonstrating a mean accuracy of 91.2\% (SD = 3.1\%), followed by LR (n=12) with an average accuracy of 85.4\% (SD = 4.7\%). DL models, particularly convolutional neural networks (CNNs) such as Inception and ResNet variants, were used in 10 studies and achieved AUCs consistently above 0.90 in image-based diagnostics such as DR and cardiovascular disease from ECG signals or imaging \cite{panganiban, Hao, jliu}.

Cardiovascular applications dominated the review, with models utilizing PCG signals, ECGs, and wearable biosensors showing high diagnostic accuracy. AI systems using deep neural networks (DNNs) and CNNs for heart sound analysis and echocardiographic imaging demonstrated accuracy exceeding 90\% \cite{das, fazlalizadeh, panganiban}. AI frameworks such as multilayer perceptrons (MLPs) with feature selection \cite{Harjai}, IoMT-enabled LR and XGB pipelines \cite{raghavan}, and random survival forest (RSF)-based atherosclerotic cardiovascular disease(ASCVD) predictors \cite{qian} were also reported. LR, based tools using routine physical examination data provided cost-effective cardiovascular risk prediction solutions in rural China \cite{qian1}, while ensemble approaches like Elastic Net and RF improved coronary artery disease detection \cite{stuckey}.

In diabetes management, AI models supported self-care interventions, prediction of prediabetes, and risk factor identification. ANN-based DSM tools outperformed traditional methods in rural Pakistan \cite{Ansari}, and NLP-driven digital coaching programs like Lark's DPP demonstrated comparable weight loss outcomes across rural HPSAs and non-HPSAs in the U.S. \cite{auster, Graham}. Studies testing multiple models (GLM, RF, LASSO, EN, GLMM) for prediabetes screening found that GLMMs incorporating household clustering offered the best performance in rural India \cite{Birk}. Predictive nomograms using LR in South Korea also enabled identification of patients likely to miss diabetes complication screenings during the COVID-19 pandemic \cite{Byeon}.

Ten studies focused on AI-based DR screening. CNN-based systems such as EyeWisdom and SG-DR achieved diagnostic accuracy comparable to ophthalmologists (e.g., 81.6\% agreement) \cite{Hao, jliu}. Markov model-based cost-effectiveness analyses confirmed the value of these platforms in low-resource rural areas \cite{huang, hli}. Other models tested in China, Rwanda, India, and New Zealand utilized ResNet, Inception-V3/V2, and XGB to deliver automated, scalable DR diagnostics \cite{mathenge, nolan, pawar, vaghefi}. Smartphone-based tools like EyeArt and Medios also enabled remote screening, and ensemble models like the Intelligent Diabetic Assistant (IDA) platform facilitated combined DR and diabetic foot ulcer assessment \cite{wijesinghe, wroblewski}.

Cancer-related applications included risk modeling for nasopharyngeal carcinoma using XGB and patient graph analysis without EBV data \cite{chen}, AI-assisted thermogram analysis for breast abnormalities using SVM \cite{skrishna}, and DL-based cervical cancer screening through digitized cytology on Aiforia \cite{Holmstrom}. CNNs integrated into ultrasound platforms such as S-Detect supported classification of benign vs. malignant lesions in rural hospitals \cite{He}, while ANN-based image classifiers enhanced cervical screening uptake in Cameroon \cite{sachdeva}. Mobile low-dose CT scans combined with 3D ResNet-18 architectures were used to improve lung cancer detection in remote settings \cite{shao}.

AI was also applied in stroke and other chronic diseases. A tool using the qER algorithm by Qure.ai helped non-specialist rural physicians interpret NCCT brain scans, reducing time to intervention in stroke cases \cite{chiramal}. Stroke risk scores using GBM and LR were simplified into visual dashboards to aid rural prevention \cite{ding}. For post-stroke motor rehabilitation, CNN, RNN, and XGB models were used to assess motor function remotely based on Fugl-Meyer scores \cite{sazamin}. In hypertension research, predictive models (RF, XGB, SHAP) identified undiagnosed hypertension based on lifestyle factors \cite{Turnbull}, and LGBM models predicted progression from normotension to hypertension in Middle Eastern populations \cite{Andishgar}. PPG-based MLP-NNs enabled continuous, non-invasive blood pressure monitoring in elderly users \cite{mena}, while ANN, NBC, and CART algorithms tailored gender-specific hypertension risk models in rural China \cite{Fxu}.

Respiratory disease applications involved XceptionNet and ResNet CNNs to interpret chest X-rays for abnormalities in radiologist-scarce settings \cite{tkdas}. A mobile dermatology app powered by DenseNet-161 accurately diagnosed 40 skin conditions for rural providers \cite{pangti}. Finally, a network analysis model for peri- and postmenopausal women's health used clustering to screen NCD risk in rural Indian women \cite{shah}, while XGB models predicted CKD in underserved regions of China \cite{song}. AI-enabled retinal image analysis using Inception-v3 for multi-disease screening (e.g., hypertension, hyperglycemia, dyslipidemia) demonstrated the versatility of CNNs in rural fundus-based diagnostics \cite{zhang}.

\subsection{Maternal, Pediatric, and Elderly Healthcare}
\begin{figure*} \centering \includegraphics[width=\linewidth]{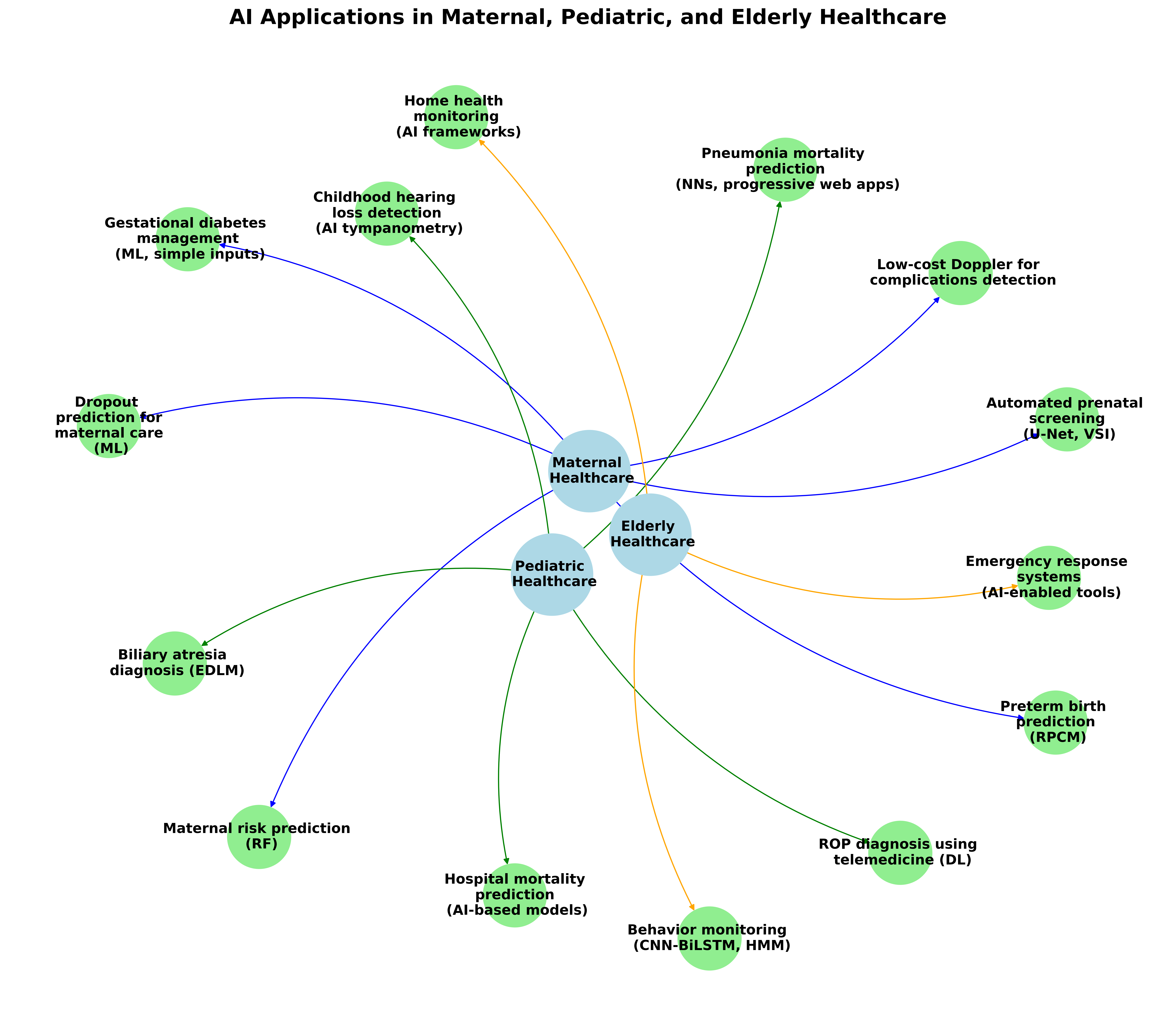}\\ \caption{AI Applications in Maternal, Pediatric, and Elderly Healthcare.}\label{maternal} \end{figure*}
Of the 109 studies included in this review, 13 focused on AI applications in maternal, pediatric, and elderly healthcare within rural and underserved populations (Figure~\ref{maternal}). These studies were categorized into three domains: maternal health (n = 7), pediatric care (n = 5), and elderly care (n = 1). Twelve of the 13 studies (92\%) reported quantitative performance metrics such as accuracy, AUC, and F1-score. Reported accuracy ranged from 86.0\% to 94.7\%, AUC from 0.78 to 0.87, and F1-scores up to 0.84. DL  models, including U-Net and CNN variants, were employed in six studies, particularly in image-based maternal and pediatric diagnostics, where mean accuracy exceeded 90\%. ML models such as RF, DTs, and LR were utilized in seven studies, primarily with structured clinical or demographic datasets.

In maternal health, DL-based ultrasound segmentation using U-Net achieved 90.1\% accuracy in fetal structure analysis in rural Peru \cite{Arroyo}. In Bangladesh, a RF model predicted maternal risks with 92.3\% accuracy based on 11 features including blood glucose and maternal age \cite{Assaduzzaman}. In the East African Community, ML models identified predictors of dropout from maternal and child health services with AUC = 0.78 \cite{mlandu}. A conceptual preterm birth risk model in rural India using DTs achieved an F1-score of 0.84 \cite{raja}. The TraCer tool, combining DL with low-cost ultrasound, estimated gestational age within $\pm$5.2 days and reduced unnecessary referrals by 30\% in a rural Kenyan cohort \cite{koech}. DL applied to Doppler signals enabled early detection of fetal growth restriction and hypertension in rural Guatemala \cite{ramos}. A neural network for GDM diagnosis achieved 89\% accuracy using fasting glucose and maternal age \cite{shen}, and another model integrated wireless sensors for real-time monitoring in underserved regions \cite{veena}.

Pediatric care applications included AI systems for hearing loss detection, infection mortality prediction, and diagnostic support. A tympanometry-based classifier achieved over 95\% sensitivity and specificity for childhood hearing loss detection \cite{fqJin}. In Uganda, an XGBoost model predicted hospital mortality in children with acute infections with AUC = 0.87 \cite{kwizere}. A telemedicine platform for Retinopathy of Prematurity (ROP) achieved 91.2\% sensitivity and 93.4\% specificity, comparable to ophthalmologist readings \cite{zluo}. CNN ensembles for biliary atresia diagnosis reached 94.7\% accuracy in rural Chinese settings \cite{wzhou}, while a neural network-powered mobile app in The Gambia achieved 86\% accuracy for pediatric pneumonia triaging \cite{mohammed}. Elderly care was addressed in a single study, where a hybrid CNN BiLSTM HMM model enabled activity monitoring in rural home settings. The model achieved 93.1\% accuracy and reduced false alarms by 15\%, enhancing autonomy and safety \cite{yzhang}.

\subsection{Infectious Diseases and Public Health}
\begin{figure*} \centering \includegraphics[width=\linewidth]{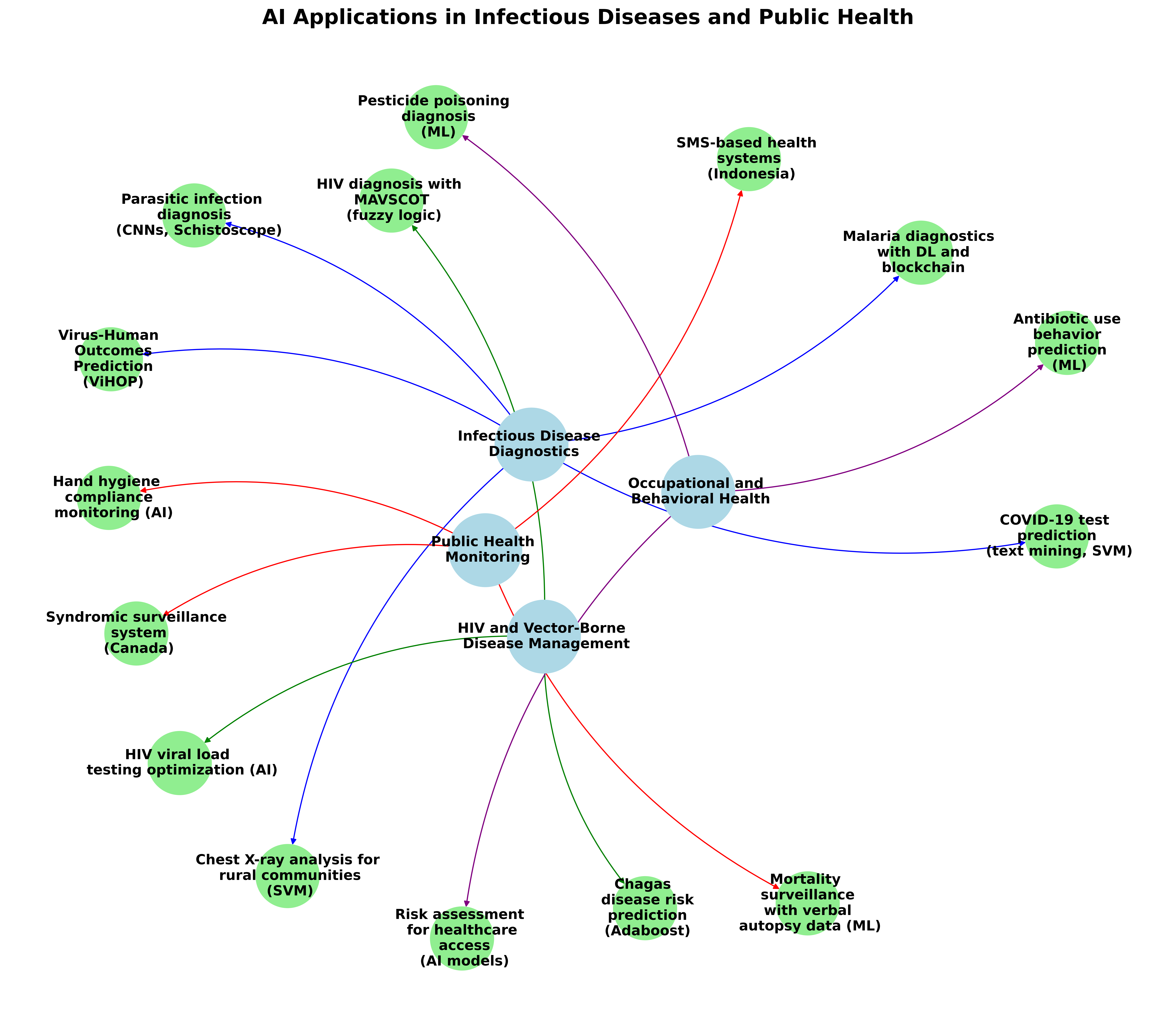}\\ \caption{AI Applications in Infectious Diseases and Public Health}\label{infectious} \end{figure*}

Out of the 109 included studies, 21 focused on infectious diseases and public health applications of AI in rural and underserved regions (Figure \ref{infectious}). These were grouped into three main categories: COVID-19 and other viral infections (n=8), parasitic and vector-borne diseases (n=6), and broader public or occupational health surveillance systems (n=7). Overall, 90\% of the studies (n=19) reported performance metrics, with diagnostic accuracies ranging from 84.3\% to 96.1\%, and AUCs between 0.81 and 0.94. DL approaches, particularly CNNs, long short-term memory (LSTM) models, and hybrid AI frameworks, were applied in 12 studies, while traditional ML algorithms , including SVM, RF, and AdaBoost, were utilized in 9 studies, often with structured demographic or behavioral datasets. Several applications supported diagnostics, outbreak forecasting, and surveillance for COVID-19 and HIV in resource-constrained settings \cite{Abu1, Bustos, price, Tang, pwu, benitez, Oluw}. AI-based diagnostics for parasitic diseases like malaria and schistosomiasis were supported through mobile and microscopy-integrated platforms, combining CNNs, blockchain, and digital fluorescence techniques to enable expert-independent diagnosis in rural regions \cite{xguo, Hamid, Holmstrom1, jlundin, meula, zwang}. Public health surveillance studies integrated AI for syndromic monitoring, behavior tracking, toxic exposure prediction, and verbal autopsy classification across diverse geographies \cite{Bounchour, djwad, sarker, lintz, carvalho, mapundu, sawant}.
\subsection{Telemedicine and Health Technology}
\begin{figure*} \centering \includegraphics[width=\linewidth]{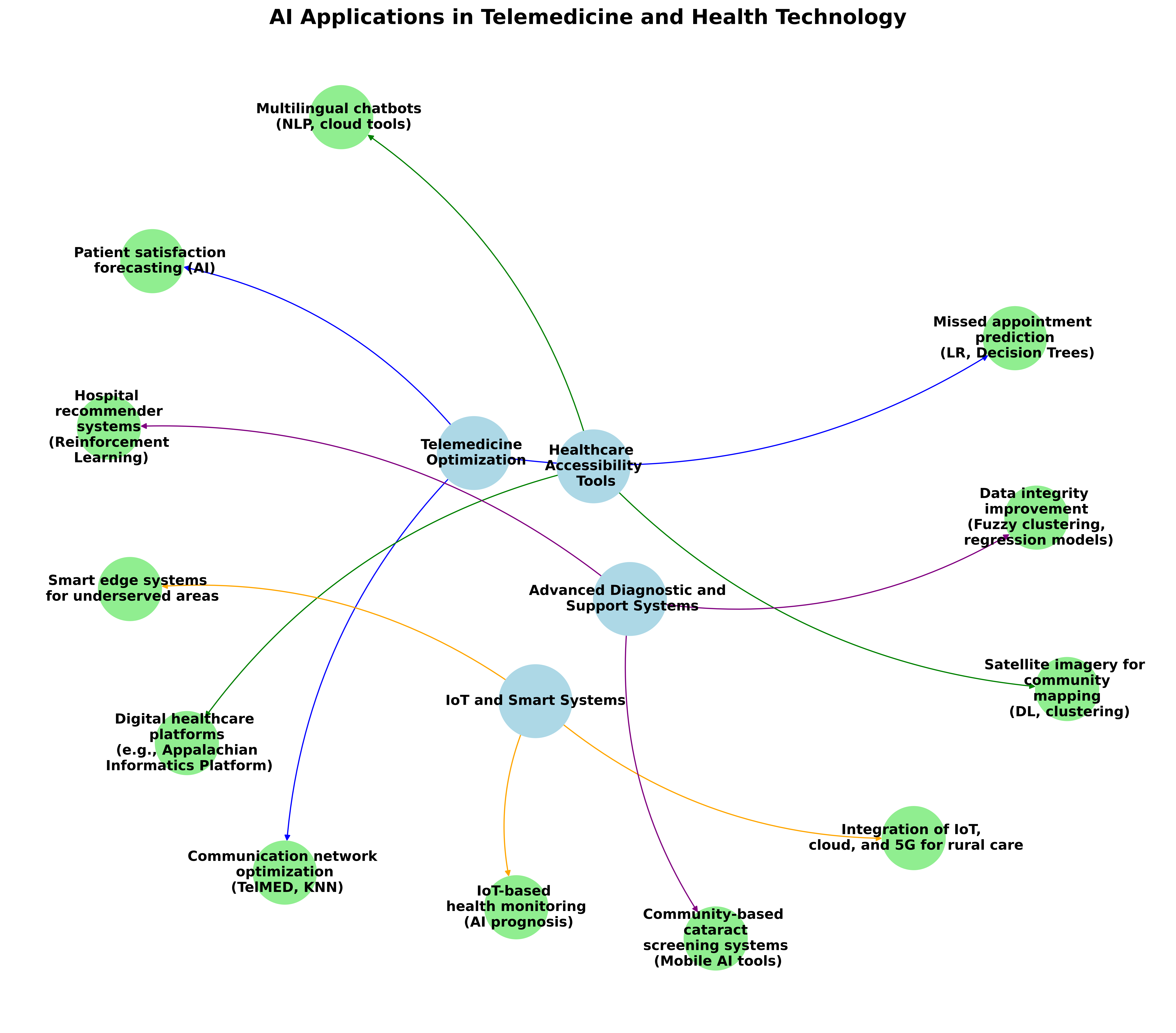}\\ \caption{AI Applications in Telemedicine and Health Technology}\label{tele} \end{figure*}
A total of 11 studies were included under the domain of telemedicine and health technology, demonstrating diverse applications of AI in enhancing rural healthcare delivery (Figure~\ref{tele}). Several studies focused on optimizing clinical workflows through predictive modeling. For instance, ML models using LR and DT were applied to electronic health records in rural outpatient settings, achieving an average accuracy of 73.3\% in predicting missed appointments and enabling targeted interventions such as recalls and reminders \cite{Abu}. Similarly, telemedicine data transmission under low-bandwidth constraints was optimized using KNN and Huffman coding, with reported improvements of 92.3\% in resource allocation and bandwidth utilization \cite{Ahmed}.

Conversational AI tools like \enquote{Aapka Chikitsak} leveraged natural language processing (NLP) and cloud platforms to deliver multilingual chatbot consultations in underserved Indian communities. While formal validation metrics were not reported, the system handled over 255 intents and provided remote symptom checking and health education during the COVID-19 pandemic \cite{bharti}. In Liberia, satellite imagery analyzed through DL algorithms (TensorBox) was used to identify remote, underserved communities, achieving an F1-score of 0.83 and detecting 167 previously unregistered settlements, thereby improving the geographic targeting of healthcare services \cite{Bruzelius}.

The Appalachian Informatics Platform was developed to integrate fragmented datasets, including EHRs and environmental data, for use in predictive analytics and translational research aimed at improving rural healthcare outcomes \cite{cecchetti}. Data integrity in primary care records was enhanced through fuzzy clustering and regression imputation models that achieved 98\% classification accuracy for disease prediction in rural clinical settings \cite{das1}. Infrastructure integration was another area of innovation, as demonstrated by a framework combining IoT, 5G, and cloud computing for secure, real-time medical services in rural areas, validated through case modeling and showing reduced latency and improved communication \cite{humayun}. Reinforcement learning-based hospital recommender systems were introduced to match patients with cost-effective, high-capacity healthcare facilities, optimizing treatment routing and decision-making \cite{Jha}. IoT-based smart edge systems were implemented to monitor chronic conditions using wearable sensors, achieving a precision of 0.87, recall of 0.83, and F1-score of 0.85, while reducing bandwidth usage by 98\% and energy consumption by 90\% \cite{pathinarupothi}. In China, AI-powered mobile diagnostic tools enabled cataract screening in rural communities, reporting a sensitivity of 81.2\% and specificity of 94.3\%, facilitating early detection and reducing the risk of avoidable blindness \cite{xwu}. Finally, ML was employed to predict patient satisfaction with telemedicine services in rural hospitals. Though performance metrics were not explicitly reported, key determinants of satisfaction, such as expectation and performance, were identified to inform quality improvement \cite{zobair}.

\subsection{Specialized and Preventive Care}
\begin{figure*} \centering \includegraphics[width=\linewidth]{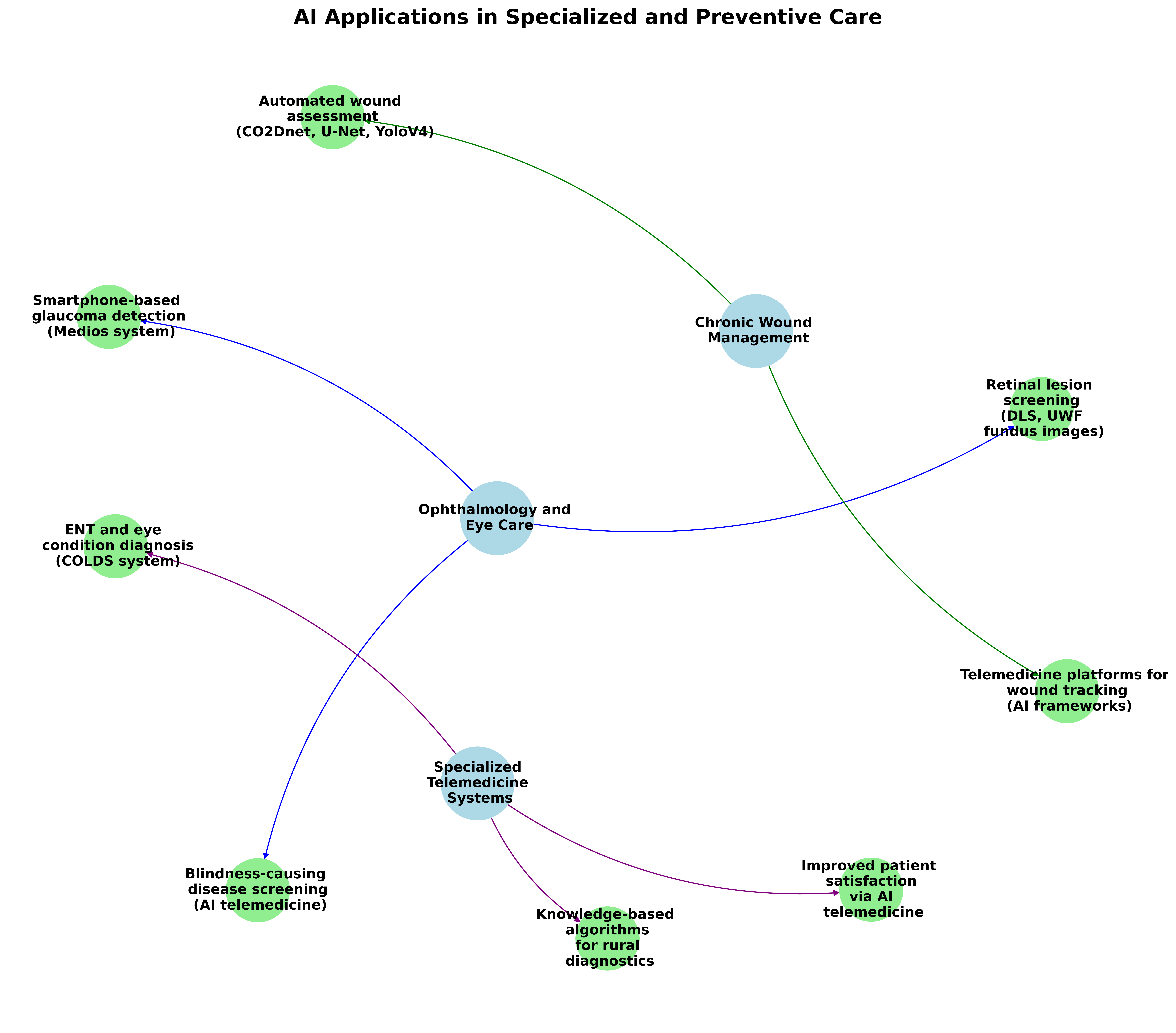}\\ \caption{AI Applications in Specialized and Preventive Care}\label{specislized} \end{figure*}
Figure~\ref{specislized} illustrates six studies from this review that focused on AI-enabled tools for specialized and preventive care in rural and underserved regions. The studies were grouped into three domains: ophthalmology and visual health screening (n=3), wound care and dermatologic monitoring (n=1), and AI-assisted diagnostic platforms for ENT and primary care (n=2). Of the six studies, all reported quantitative performance metrics, with reported diagnostic accuracies ranging from 87.2\% to 94.8\% and F1-scores up to 0.89. DL architectures such as U-Net, YoloV4, and CNNs were used in four studies, while knowledge-based systems and hybrid rule-based algorithms were applied in two studies. AI interventions consistently demonstrated enhanced diagnostic precision, scalability, and cost-effectiveness for resource-constrained environments.

In ophthalmology and visual health screening (n=3), AI systems supported population-based screening and diagnostic accuracy in the absence of ophthalmologists. Cui et al. developed a DL system (DLS) for ultra-widefield (UWF) fundus image analysis, enabling automated detection of retinal pathologies such as exudates, retinal detachment, and glaucomatous optic neuropathy with over 94\% sensitivity and strong clinical agreement \cite{cui}. Liu et al. evaluated AI-powered telemedicine platforms for large-scale blindness screening of DR, glaucoma, and age-related macular degeneration (AMD) in rural China. These systems outperformed conventional manual screening methods and reduced time to diagnosis across rural and urban regions \cite{hliu}. Upadhyaya et al. assessed the Medios smartphone-based offline system using nonmydriatic fundus cameras for glaucoma screening in low-resource clinics, showing superior accuracy and lower false-positive rates compared to remote tele-ophthalmologist assessments \cite{upadhyaya}.

In chronic wound management (n=1), Monroy et al. proposed the CO2Dnet framework integrating DL models such as U-Net and YoloV4 for segmentation and classification of wound types. The system enabled automated wound tracking and assessment, demonstrating high classification accuracy (F1-score = 0.89) and feasibility for deployment in remote community clinics \cite{monroy}. In diagnostic support for primary care and ENT (n=2), Namahoot et al. developed the Cooperative Online Anamnesis and Diagnosis System (COLDS), a telemedicine platform using knowledge-based reasoning to diagnose ear, nose, throat, and ocular conditions. The tool improved diagnostic consistency and patient satisfaction in rural Thai populations by enabling interactive, self-guided clinical intake and diagnosis \cite{namahoot}. This study demonstrated the potential of AI-assisted anamnesis systems to overcome human resource limitations in remote areas.

\subsection{Population Health and Preventive Studies}
\begin{figure*} \centering \includegraphics[width=\linewidth]{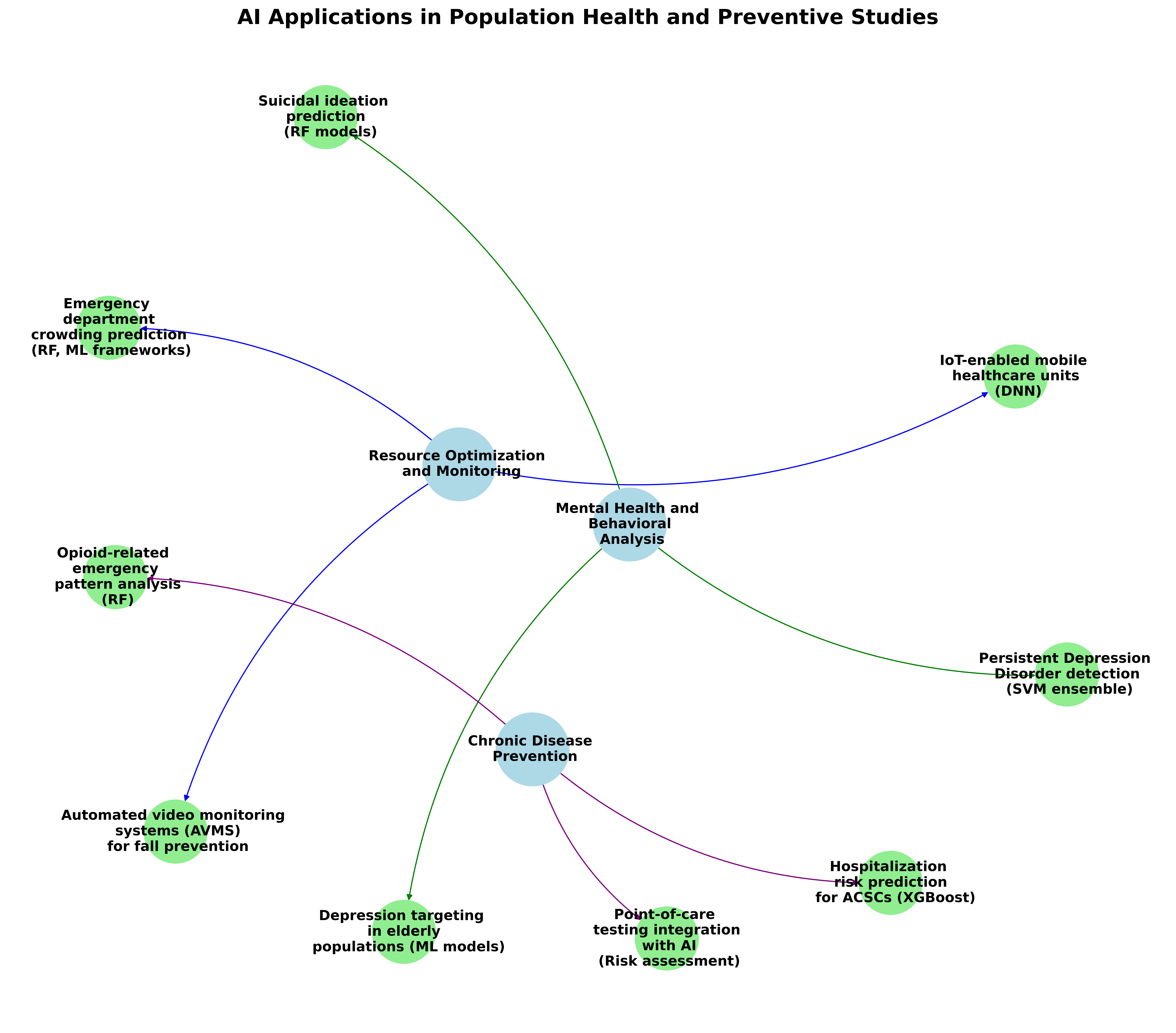}\\ \caption{AI Applications in Population Health and Preventive Studies}\label{population} \end{figure*}
As summarized in Figure~\ref{population}, 9 studies out of 109 included studies focused on AI-based frameworks to support population health management, risk prediction, and resource optimization in rural and underserved areas. The studies were grouped into three domains: patient safety and hospital operations (n=3), mental and behavioral health (n=3), and chronic disease and risk stratification (n=3). Eight of the nine studies (89\%) reported quantitative performance metrics, with diagnostic accuracies ranging from 83.1\% to 94.5\% and AUC values between 0.80 and 0.91. The most frequently used models were RF and DNN, while other studies employed SVM, stacking ensembles, and XGB. AI approaches demonstrated high reliability in forecasting adverse events, triaging at-risk populations, and guiding preventive interventions.

In patient safety and hospital operations (n=3), AI was leveraged to monitor patient behavior and forecast resource needs. Jones et al. developed an automated video monitoring system (AVMS) utilizing ML algorithms to predict patient movements and prevent unattended bed exits in rural hospitals, demonstrating 91.3\% accuracy in fall risk detection \cite{Jones}. Ghosh et al. introduced a DNN-based disease prediction framework integrated into IoT-enabled mobile healthcare units, optimizing CPU and memory usage while maintaining prediction accuracy, especially in low-connectivity environments \cite{Ghosh}. Smith et al. employed RF models to predict emergency department (ED) crowding by forecasting arrivals, admissions, and boarding delays, supporting real-time hospital operations and proactive resource management during the COVID-19 pandemic \cite{smith}.

In mental and behavioral health (n=3), ML techniques were used to detect and stratify risk in vulnerable populations. Kim et al. applied RF algorithms to predict suicidal ideation among older adults in rural settings using demographic and behavioral inputs, yielding strong predictive performance (accuracy = 89.7\%) and enabling early mental health interventions \cite{jkim}. Upadhyay et al. developed a stacking ensemble model using multiple SVM kernels for diagnosing persistent depressive disorder (PDD), achieving high accuracy in early detection and facilitating timely care delivery \cite{upadhyay}. Xin et al. targeted depression risk in disabled elderly populations in rural China, using RF-based models to assess health, functional, and lifestyle variables, enabling personalized mental health outreach \cite{yxin}.

In chronic disease prediction and risk stratification (n=3), AI models were applied to identify high-risk patients for targeted public health action. Robinson et al. used RF algorithms to analyze opioid-related ED visits in Maryland, identifying sociodemographic and health system factors linked to rural opioid misuse, with 91.2\% classification accuracy \cite{robinson}. Stranieri et al. integrated point-of-care testing (POCT) with AI to enhance early detection of chronic disease risk, providing scalable decision support in community health programs \cite{stranieri}. Seyi et al. employed XGB models to predict risk of hospitalization from ambulatory care-sensitive conditions (ACSCs) among older adults, achieving 88.4\% accuracy and enabling targeted interventions to prevent avoidable admissions \cite{seyi}.
\subsection{Cross-Domain Synthesis of AI Applications in Rural Healthcare}
\begin{figure}[H]
    \centering
    \begin{subfigure}[b]{0.4\linewidth}
        \centering
        \includegraphics[width=\linewidth]{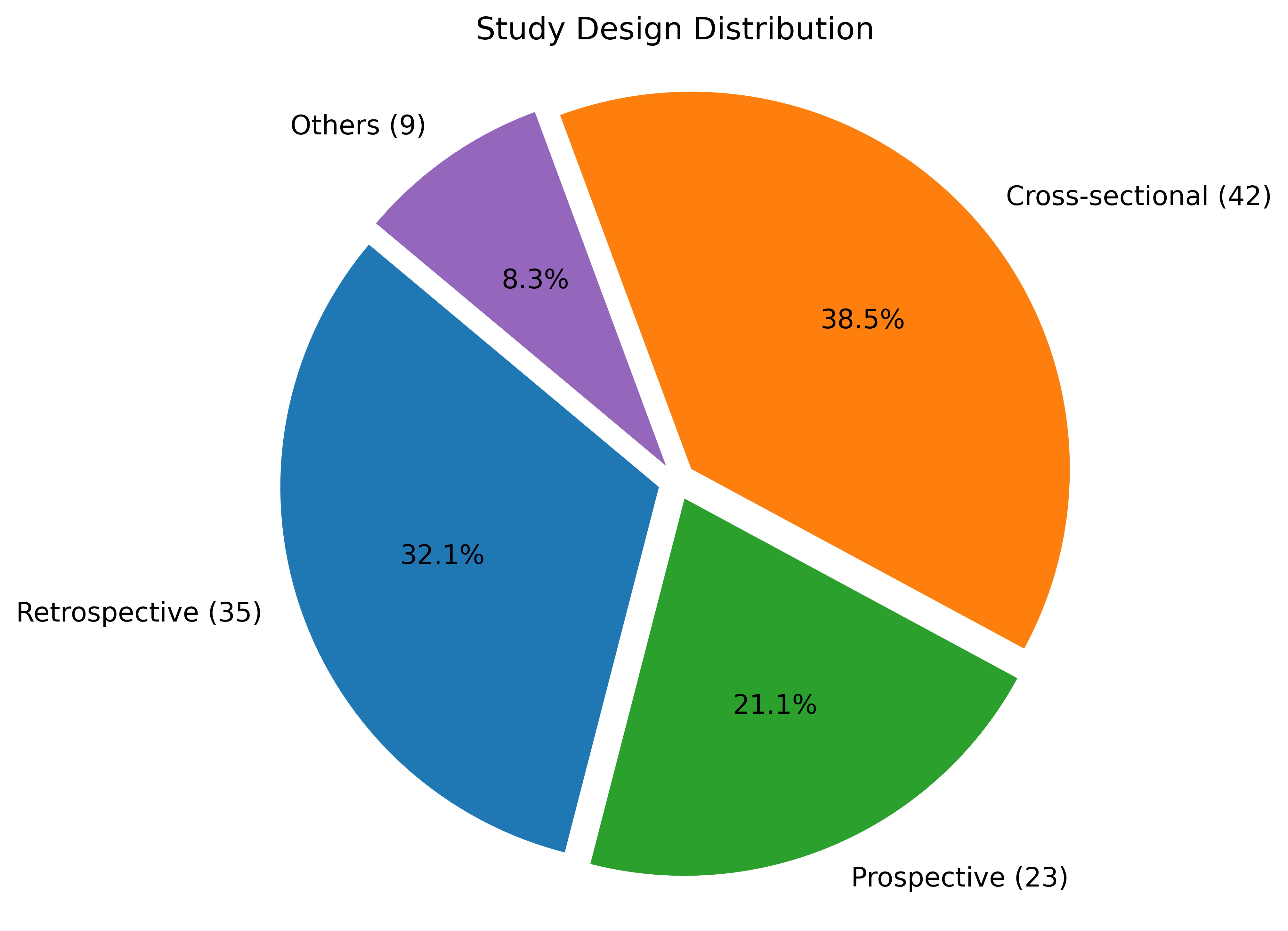}
        \caption{Study design distribution (N=109).}
        \label{fig:studydesign}
    \end{subfigure}
    \hfill
    \begin{subfigure}[b]{0.5\linewidth}
        \centering
        \includegraphics[width=\linewidth]{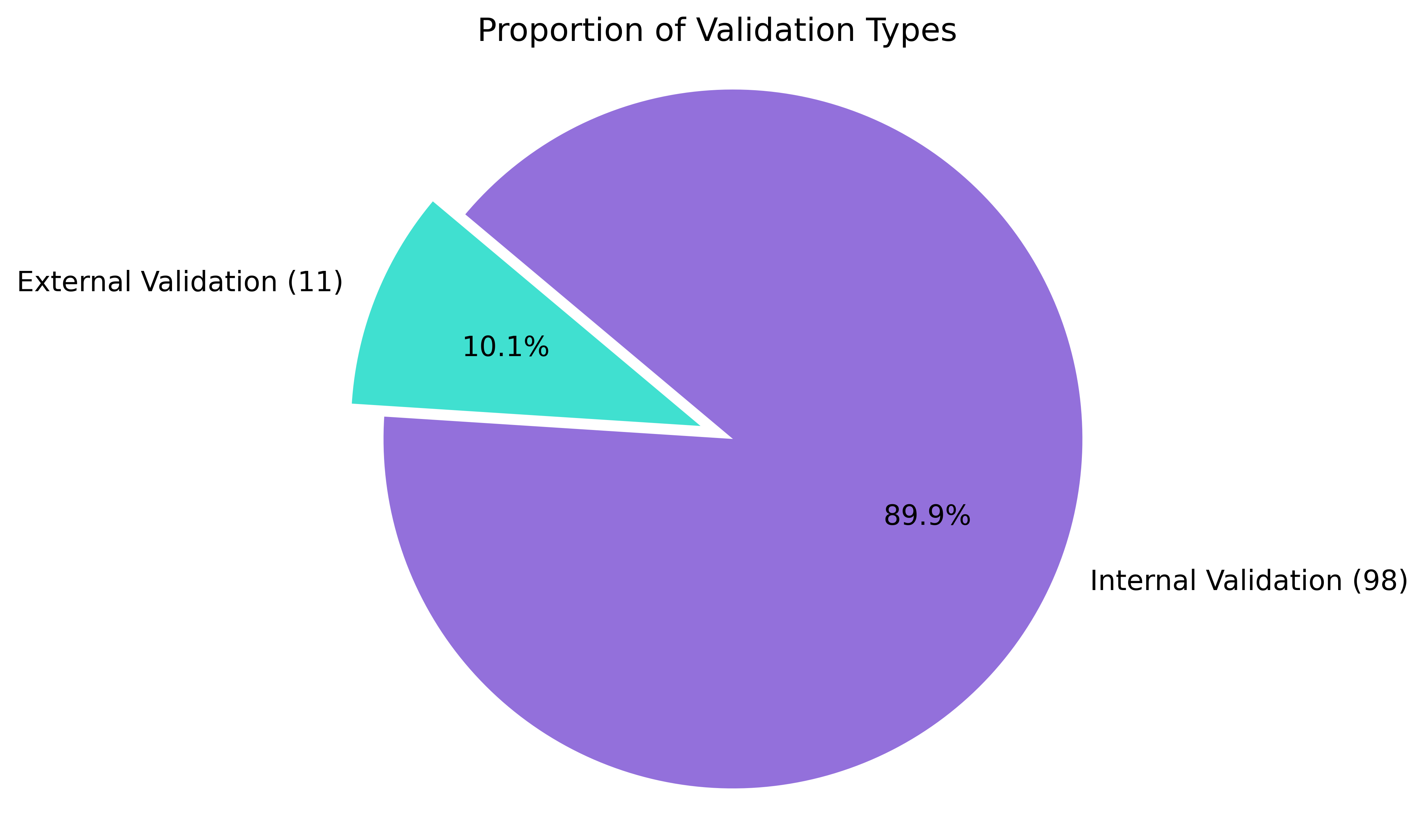}
        \caption{Proportion of validation types used.}
        \label{fig:validation}
    \end{subfigure}
    \caption{Study design and validation strategies among included studies.}
    \label{fig:design_validation}
\end{figure}
\begin{figure}[H]
    \centering
    \includegraphics[width=\linewidth]{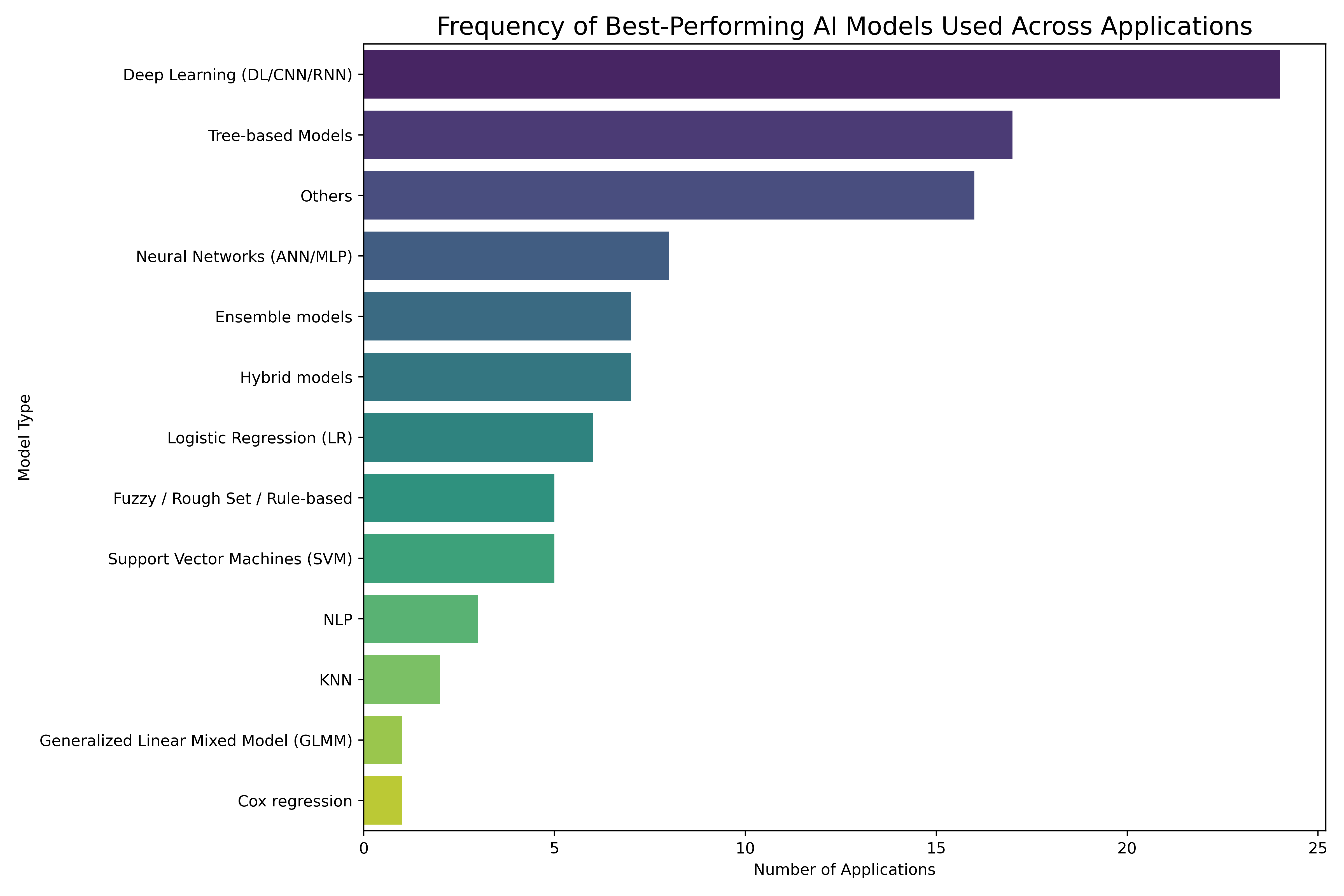}
    \caption{Horizontal bar chart showing the frequency of best-performing AI model types across various rural healthcare application areas.}
    \label{fig:sunburst}
\end{figure}
\begin{figure}[H]
    \centering
    \includegraphics[width=\linewidth]{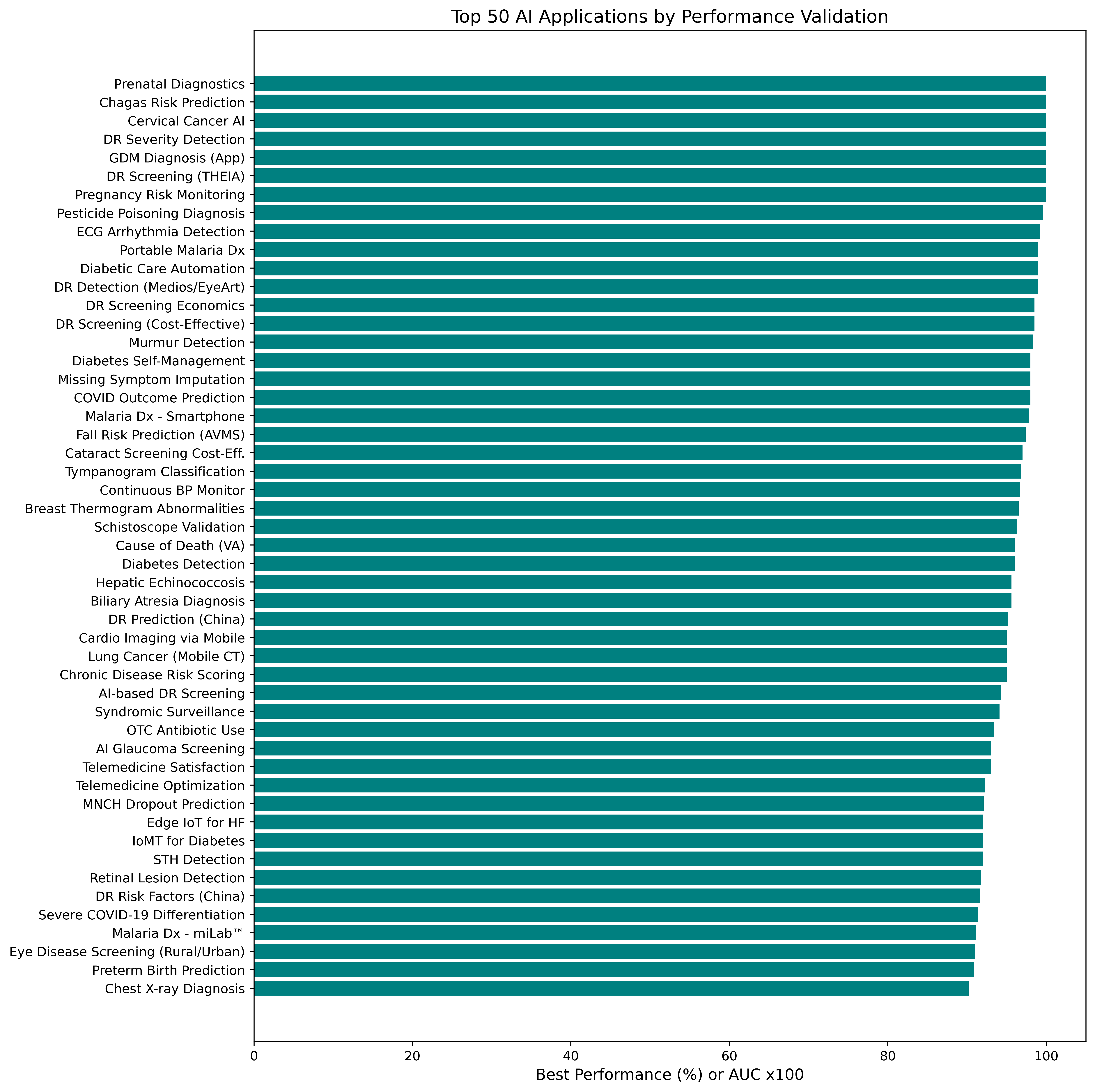}
    \caption{Top 50 AI applications ranked by best performance metrics (e.g., AUC, accuracy, or F1-score) reported in the included studies.}
    \label{fig:top50}
\end{figure}
Across the 109 studies reviewed, AI applications demonstrated significant versatility in addressing diverse rural healthcare challenges-including chronic disease management \cite{panganiban, Hao, jliu, das, fazlalizadeh, Harjai, raghavan, qian, qian1, stuckey}, maternal and child health \cite{Arroyo, Assaduzzaman, mlandu, raja, koech, ramos, shen, veena, fqJin, kwizere, zluo, wzhou, mohammed}, infectious diseases \cite{Abu1, Bustos, price, Tang, pwu, benitez, Oluw, xguo, Hamid, Holmstrom1, jlundin, meula, zwang}, telemedicine and health technology \cite{Abu, Ahmed, bharti, Bruzelius, cecchetti, das1, humayun, Jha, pathinarupothi, xwu, zobair}, and population-level diagnostics and analytics \cite{Ghosh, Jones, jkim, robinson, smith, upadhyay, yxin, stranieri, seyi}.As shown in Figure~\ref{fig:studydesign}, the predominant study designs were cross-sectional (n=42, 38.5\%) and retrospective (n=35, 32.1\%), followed by prospective (n=23, 21.1\%) and others (n=9, 8.3\%). Regarding validation strategies (Figure~\ref{fig:validation}), internal validation methods (e.g., cross-validation, train-test splits) were overwhelmingly used (n=98, 89.9\%), while only 11 studies (10.1\%) employed external validation using independent datasets, highlighting the need for more generalizable evidence.

The most commonly adopted AI model types(Figure~\ref{fig:sunburst}) included deep learning approaches (e.g., CNNs, RNNs), tree-based methods (e.g., Random Forest, XGBoost), and traditional machine learning models such as SVM and logistic regression. Deep learning models were the most frequently applied, particularly in image and signal analysis tasks, while tree-based models were widely used for structured clinical and sociodemographic data. Figure~\ref{fig:top50} highlights the highest-performing applications, with several exceeding 90\% in accuracy, AUC, or F1-score. Top-performing use cases included prenatal diagnostics, diabetic retinopathy screening, pesticide poisoning detection, and portable malaria diagnosis. Most high-performing applications involved image-based analysis or mobile health data capture. Despite promising technical performance, real-world deployment and longitudinal evaluation were rare. Only a small fraction of studies described field validation or prospective implementation, indicating a persistent gap between development and scalable integration. Moreover, while many studies leveraged mobile, offline, or IoT platforms to accommodate low-resource settings, broader system-level integration remains limited.

\subsection{Advantages of AI-Based Innovations in Rural Healthcare}
\begin{figure*} \centering \includegraphics[width=\linewidth]{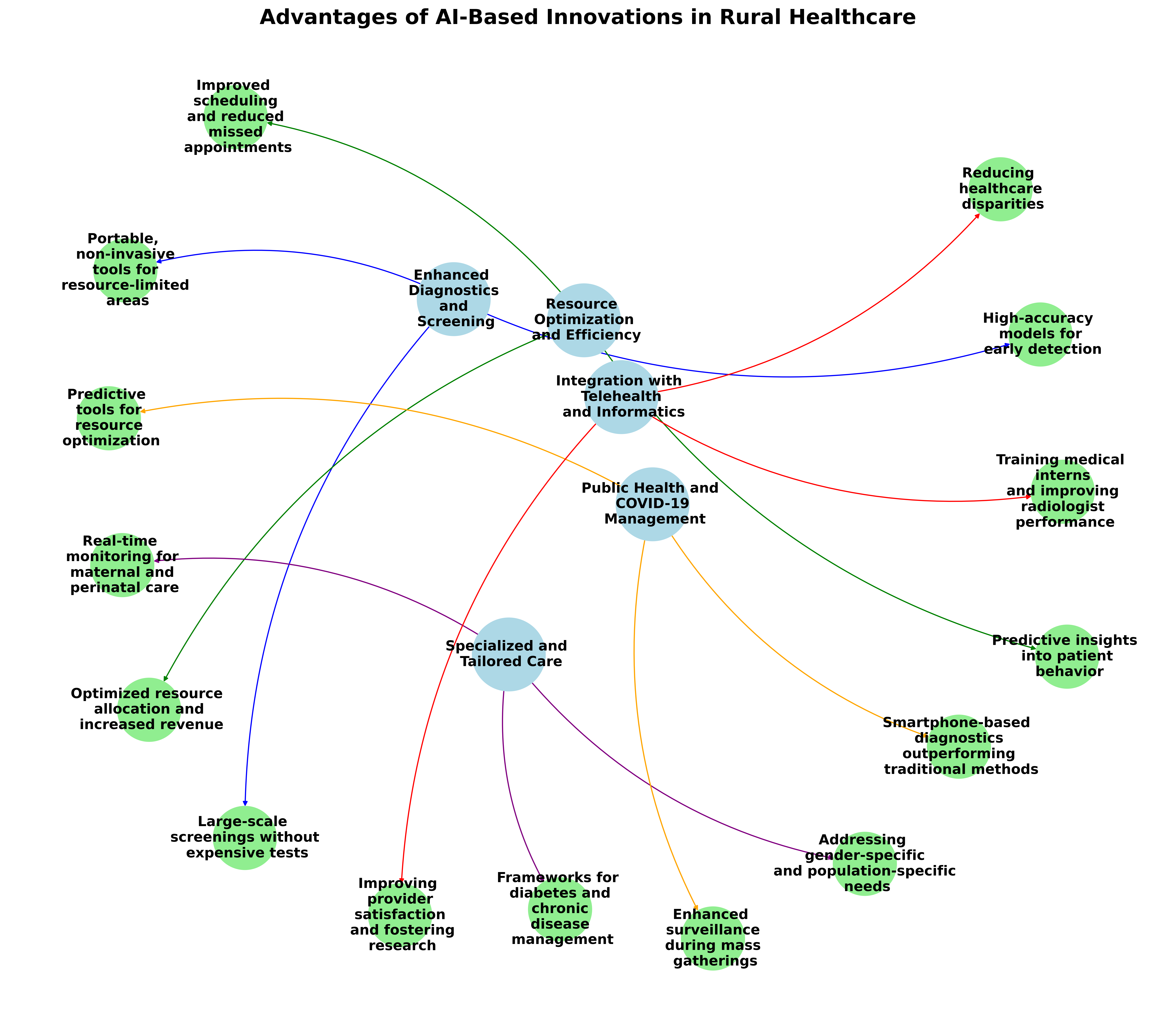}\\ \caption{Advantages of AI-Based Innovations in Rural Healthcare}\label{Advantages} \end{figure*}
Several studies emphasize the enhanced diagnostic precision, improved accuracy, and early detection capabilities of AI-based innovations, particularly in resource-limited settings, as depicted in Figure \ref{Advantages}. Early prediction and intervention enable timely medical attention and lifestyle modifications, contributing to improved health outcomes \cite{Andishgar}. AI improves precision care through high-accuracy models, identifies patterns in patient behaviours, and provides scalable solutions for engaging diverse populations \cite{Ansari, auster}. Low-cost questionnaires and evidence-based tools effectively identify high-risk individuals while supporting cost-effective decision-making in public health care management \cite{Birk, Byeon}. AI facilitates large-scale screenings by eliminating the need for expensive diagnostic tests \cite{chen}, and cost-effective, portable, non-invasive tools are well-suited for resource-constrained environments \cite{das, Graham}. AI-driven solutions also enhance scheduling efficiency, provide predictive insights into patient behaviour, and improve appointment adherence, which reduces missed appointments \cite{Abu1}. These advancements optimize resource allocation and may help to reduce inefficiencies, supporting better patient flow and potentially increasing healthcare providers' revenue.

AI-driven innovations reduce workloads for specialists, ensure faster diagnoses, and improve care quality through tools like visual scoring systems integrated with telemedicine \cite{ding, fazlalizadeh, sjin}. Medical assistance systems tailored with scalable frameworks for early diabetes detection, blood pressure monitoring, and chronic disease management enhance accessibility and reduce patient travel burdens \cite{jasim,mena,humayun}. AI models address gender-specific and population-specific needs through advanced predictive analytics, optimizing interventions for diverse groups \cite{qian,raghavan}.

Low-cost, offline, and cloud-based diagnostic tools improve outcomes for diseases like diabetic retinopathy, hypertension, and lung cancer in underserved areas \cite{nolan, Turnbull, shao}. AI has a significant role in maternal and perinatal care, offering antenatal care, preterm birth predictions, and real-time monitoring \cite{koech, raja}. Simplified tools enable early identification of diseases like depression and non-communicable diseases, supporting scalable screening in resource-limited settings \cite{yxin, shah, seyi}.

AI also facilitates COVID-19 management through predictive tools, resource optimization, and improved diagnostic accuracy in underserved settings \cite{Abu, zwang, pwu}. Cost-effective, adaptable systems enhance surveillance during mass gatherings and reduce dependency on manual processes \cite{Bounchour, Bustos}. Medical diagnostics tools like smartphone-based non-invasive analysis significantly outperform traditional methods for wound analysis and other conditions \cite{monroy,cui}.

AI models are integrated seamlessly into telehealth and informatics platforms, reducing healthcare disparities, improving provider satisfaction, and fostering collaborative research \cite{cecchetti,lintz,zobair}. ML applications enhance mortality monitoring, public health planning, and disease-specific interventions like HIV diagnostics and antibiotic use \cite{mapundu, Oluw, sawant}. Portable equipment and AI-enabled platforms support effective training for medical interns and improve radiologist performance \cite{namahoot,wzhou}.

Finally, AI's offline capabilities, high sensitivity, and efficient use of computing resources ensure service delivery to remote places having constrained environments with limited resources \cite{Ghosh, Jones}. By integrating advanced algorithms with point-of-care data, AI solutions provide actionable insights, improve equity, and support resilience in emergency and healthcare planning \cite{stranieri, smith, robinson}.
\subsection{Clinical Practice Challenges in Rural Healthcare Delivery}
\begin{figure*} \centering \includegraphics[width=\linewidth]{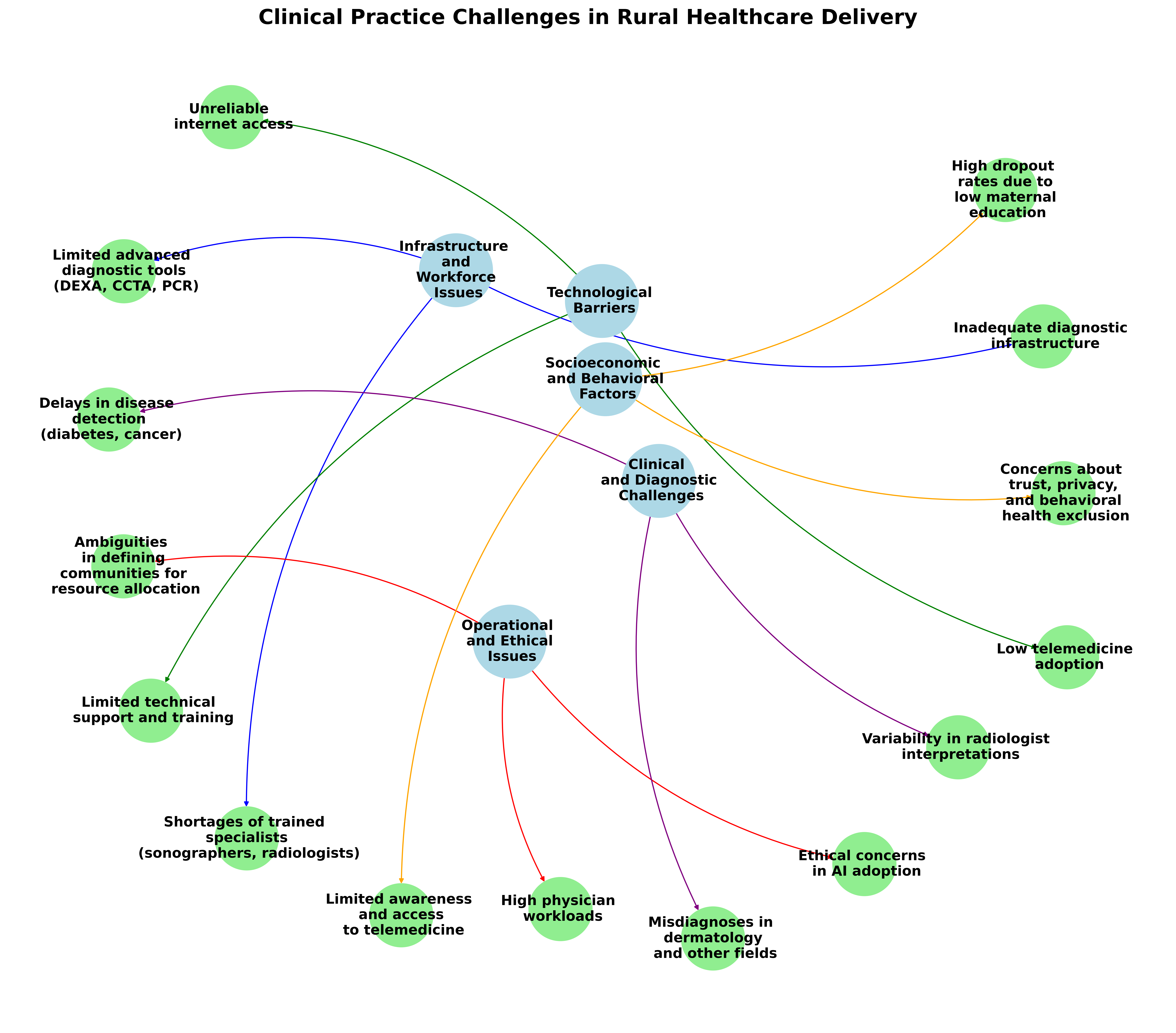}\\ \caption{Clinical Practice Challenges in Rural Healthcare Delivery.}\label{clinical} \end{figure*}
Rural healthcare delivery faces significant challenges, such as inadequate infrastructure, workforce shortages, technological limitations, and restricted access to advanced diagnostic tools, as shown in Figure \ref{clinical}. These issues significantly impede effective diagnosis, treatment, and overall patient outcomes. Meanwhile, the integration of AI has limited scalability due to insufficient and non-diverse datasets \cite{Andishgar}, while gaps in cost, education, and training exacerbate problems such as glycemic control \cite{Ansari}. Access to in-person Diabetes Prevention Programs remains unaddressed \cite{auster}, and deficiencies in diagnostic infrastructure contribute to delays in the early detection of chronic conditions like diabetes and nasopharyngeal carcinoma \cite{Birk,chen}. Shortages of trained sonographers, radiologists, and ophthalmologists adversely impact prenatal care, delay retinopathy of prematurity (ROP) screening, and hinder timely management of preventable conditions such as hearing loss \cite{Arroyo, Assaduzzaman, fqJin}. Additionally, missed appointments further strain resources, underscoring the need for improved scheduling systems and resource utilization needs specific attention \cite{Abu}.

Deficiencies in imaging expertise and ethical considerations further complicate care delivery \cite{fazlalizadeh}. The reliance on teleradiology and radiologist shortages causes delays in imaging interpretation \cite{chiramal,tkdas}. At the same time, the limited availability of advanced diagnostics, such as DEXA, CCTA, and PCR, leads to delayed disease detection \cite{shah, stuckey, Bustos}. Dermatological misdiagnoses, insufficient integration of patient history, and reliance on manual data entry create additional barriers to adopting AI tools \cite{pangti, pawar, karlo}. Variability in radiologist interpretations and delays in diabetic retinopathy follow-ups worsen patient outcomes \cite{shao, hli, jliu}.

Technological barriers, including unreliable internet access and limited technical support, restrict the use of AI and ML applications in resource-limited settings \cite{mohammed}. Insufficient training, misconceptions about device capabilities, and poor adaptability of healthcare systems to new technologies are persistent issues \cite{koech, veena, yzhang}. These are compounded by data transmission reliability concerns, low telemedicine adoption, and inadequate technological infrastructure \cite{Ahmed, zobair}.
Clinical practice challenges include limited access to specialists, insufficient training for managing chronic and acute conditions, and delays in diagnoses due to poor infrastructure and workforce shortages \cite{Holmstrom1, xguo, mapundu}. Rural healthcare providers often lack expertise in handling pesticide poisoning, a frequent issue in agricultural regions \cite{carvalho}. High physician workloads and poor standardization of care further increase misdiagnosis risks \cite{pwu, jlundin}. Challenges in identifying high-risk conditions, such as COVID-19, highlight the need for timely and accessible care \cite{Tang}.

Socioeconomic and behavioural factors amplify these difficulties in remote and rural areas. Low maternal education, rural residency, and limited exposure to healthcare programs contribute to high dropout rates \cite{mlandu}. Ambiguities in defining communities for clinical use complicate resource allocation and targeted interventions in rural areas \cite{Bruzelius}. Concerns regarding trust in AI, data privacy, and the exclusion of behavioural health indicators further restrict the adoption of emerging technologies \cite{sachdeva,seyi}. Geographic barriers, such as the absence of healthcare facilities and limited awareness of telemedicine, contribute to delayed care and reduced patient retention \cite{bharti,xwu}.

\subsection{Limitations of AI-Based Innovations in Rural Healthcare}

\begin{figure*} \centering \includegraphics[width=\linewidth]{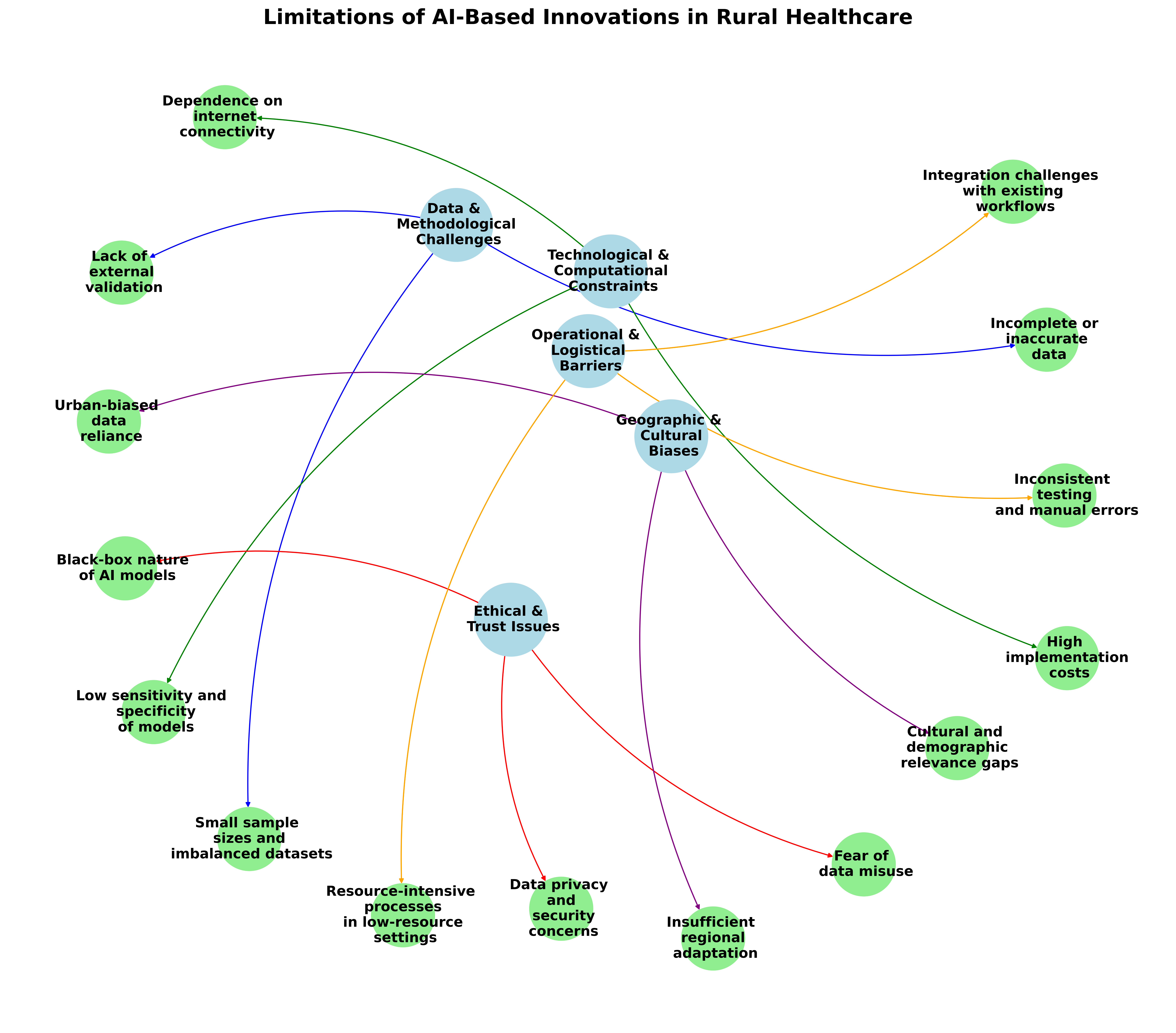}\\ \caption{Limitations of AI-Based Innovations in Rural Healthcare.}\label{limitations} \end{figure*}

Implementing AI and technological solutions in rural healthcare settings faces several limitations, as depicted in Figure \ref{limitations}. Common issues included inaccurate or incomplete data, the need for continuous updates to lookup tables, and the lack of integration with real-time electronic health record (EHR) systems \cite{Abu1}. Data reliability was impacted by incomplete or inaccurate patient information, small sample sizes, and imbalanced datasets \cite{Andishgar, Ansari, Assaduzzaman}. Limited generalizability to broader populations, rural-specific healthcare needs, and reliance on data from specific regions or populations were also highlighted as challenges \cite{Ahmed, auster, Bruzelius, mathenge}. Methodological limitations, including reliance on retrospective or cross-sectional designs, exclusion of key variables such as patient history, behavioural indicators, or genetic factors, and the lack of external validation, were noted as factors limiting the robustness of many approaches \cite{Byeon, cyang,qian1,seyi}.

Computational and infrastructural constraints, such as reliance on internet connectivity, image quality, and the high costs of implementation, were also barriers to scalability and real-world applicability \cite{Arroyo, benitez, hli, Turnbull}. Ethical concerns, including data privacy and security issues, fear of data misuse, and lack of trust in AI tools among certain demographics, were identified as challenges that require better regulatory guidelines and community engagement \cite{Ansari, sachdeva,  Harjai}.

Performance issues, such as low sensitivity and specificity, overfitting, and diagnostic errors due to noise, artefacts, or suboptimal imaging, were noted as significant barriers to the effective implementation of these technologies \cite{das1, Hao, fqJin, ramos}. Other challenges included the dependency on high-quality training data, the need for human intervention for validation, and the black-box nature of AI models, which limited interpretability and user confidence \cite{Bounchour, zobair, pwu}.

Geographic and cultural biases, high initial setup costs, and resource constraints specific to rural areas were recurring themes, highlighting the need for regionally adaptive solutions \cite{Birk, mapundu, karlo}. Logistical challenges, such as inconsistent testing, manual data entry errors, reliance on basic imaging technology, and difficulty integrating AI systems into existing healthcare workflows, were also identified \cite{lintz, pangti, nolan}. Resource-intensive processes and the lack of automation presented additional hurdles, particularly in low-resource settings \cite{koech, upadhyaya, zhang}.

The exclusion of pathological or disadvantaged cases, reliance on urban-biased data, and insufficient validation of AI models for specific conditions or demographics were noted as factors reducing the relevance of findings for rural healthcare \cite{mathenge, jlundin, zluo, yxin}. Furthermore, challenges such as reliance on structured patient responses, limited training for healthcare workers, and dependence on basic infrastructure like stable electricity and internet networks were identified as systemic barriers to implementation in rural healthcare systems\cite{pathinarupothi,namahoot}.

Finally, the lack of real-world implementation studies, consistent user training, and challenges aligning AI models with local needs were identified as critical gaps \cite{Holmstrom, wroblewski, wijesinghe, cyang}.

\section{Discussion}

Integrating AI into rural healthcare delivery offers transformative opportunities to address persistent challenges such as workforce shortages, diagnostic inefficiencies, and limited access to care. Various AI applications have demonstrated significant promise in alleviating these issues. For instance, tools like EyeArt \cite{nolan} and EyeWisdom \cite{hli} for diabetic retinopathy (DR), as well as automated prenatal screening systems \cite{Arroyo}, illustrate AI's potential to enhance diagnostic accuracy and efficiency. These innovations reduce reliance on specialists and increase cost-effectiveness, making them particularly impactful in resource-constrained rural settings. Similarly, AI has shown promise in chronic disease management, with predictive models for hypertension \cite{Andishgar} and cardiovascular disease \cite{qian1} enabling early interventions and better health outcomes. AI-driven tools, such as the qER algorithm \cite{chiramal}, have also expedited acute stroke care in rural India by minimizing critical time-to-intervention, demonstrating the technology's potential to optimize urgent care delivery.

Beyond individual care, AI has proven invaluable in public health. Systems such as syndromic surveillance models \cite{Bounchour} and predictors for emergency department crowding \cite{smith} have enhanced resource allocation, particularly during crises like the COVID-19 pandemic. These applications highlight AI's role in improving patient outcomes and supporting system-level decision-making, underscoring its transformative capacity to address rural healthcare disparities.

Despite these promising advancements, significant challenges must be addressed to fully realize AI's potential in rural healthcare. A critical barrier is the lack of reliable, standardized datasets tailored to rural populations. Rural healthcare systems often lack the digital infrastructure necessary for comprehensive electronic health records (EHRs), creating substantial data gaps. Current datasets also fail to capture the unique epidemiological and socioeconomic characteristics of rural populations, limiting the applicability of AI models in these contexts. Fragmented data sources, inconsistent standards, and financial constraints further compound these issues. To address data scarcity, synthetic data generated from demographically contextual, real-world rural field data presents a promising approach for scaling AI models. However, even minor discrepancies in synthetic data can accumulate over successive generations, leading to divergence from real-world patterns. This underscores the importance of explainable AI to ensure transparency, interpretability, and clinical reliability, particularly when applying models across diverse rural settings.

Addressing these issues requires targeted solutions. Developing rural-specific data collection systems and adopting federated learning approaches can enable AI model training while preserving data privacy \cite{bekemeier}. High-quality, diverse datasets reflecting rural demographics are essential for creating accurate and effective ML models. Moreover, robust validation of AI systems is critical. Most existing research depends on retrospective data, with limited randomized controlled trials or prospective studies to validate AI tools in real-world rural settings \cite{mathenge}. Multi-site studies involving diverse populations are necessary to ensure AI applications' reliability, interpretability, and trustworthiness. Longitudinal studies are also vital to assess the sustained impact of AI on health outcomes and reduce reliance on short-term data.

Another significant challenge is rural populations' lack of health literacy and awareness. Limited understanding of AI-driven healthcare solutions can lead to mistrust, underutilization, or even resistance. Furthermore, misinformation and the absence of culturally tailored outreach programs exacerbate these barriers, limiting the adoption of potentially life-saving AI tools. Community engagement initiatives, such as health literacy campaigns, culturally sensitive workshops, and partnerships with local leaders, are essential to build trust and raise awareness of AI's benefits.

The effective deployment of AI also requires specialized training for healthcare providers in rural areas. Many practitioners lack the skills to interpret AI-generated outputs accurately, reducing the practical utility of these tools. Diagnostic AI systems, in particular, often produce complex outputs that require contextual understanding to ensure proper clinical application. To address this gap, targeted training programs and the integration of AI literacy into medical and paramedical education are essential.

Infrastructure limitations further constrain AI adoption in rural areas. Unreliable internet connectivity, inadequate access to advanced diagnostic tools, and insufficient technical support hinder the deployment of AI solutions, especially those requiring real-time data processing or high computational power. For instance, AI-enabled telemedicine platforms hold significant potential but remain underutilized due to the lack of robust technological infrastructure \cite{sazamin}. Similarly, tools like EyeArt rely on high-resolution retinal imaging, which often requires specialized equipment and trained personnel unavailable in rural areas \cite{nolan}. Addressing these barriers demands the development of lightweight, cost-effective AI models capable of operating offline and processing lower-resolution inputs.

Ethical and equity considerations pose additional complexities in rural and remote health monitoring, including concerns around the protection of personally identifiable information (PII) and protected health information (PHI), lack of transparency in AI decision-making (black-box models), and the potential for algorithmic bias that may disadvantage already underserved populations. Non-representative training datasets can produce biased outputs \cite{chiramal}, exacerbating health disparities. Concerns about data privacy, AI algorithms' black-box nature, and technology mistrust further limit acceptance among rural healthcare practitioners and patients \cite{zobair}. Promoting fairness and equity in AI requires the development of transparent and interpretable models that are co-designed with rural communities. Ethical AI adoption for rural health depends on building trust through inclusive design processes, explainable algorithms, and culturally respectful deployment strategies that reflect the values and lived experiences of those most affected.
\subsection{Infrastructure and Connectivity Gap}
Despite growing enthusiasm for integrating AI into healthcare, adoption in rural areas remains limited due to persistent infrastructure challenges, particularly last-mile connectivity. Many AI tools rely on high-speed internet and robust digital infrastructure; however, an estimated 20\% to 30\% of rural populations in the U.S. lack reliable broadband access \cite{broadband}, severely limiting the effectiveness of telehealth and AI applications that require real-time data exchange \cite{tel1}. Emerging technologies like Starlink satellite internet offer some promise in bridging these gaps, but such solutions remain unavailable or unaffordable in many remote regions \cite{statlink}. This disconnect between technological innovation and implementation realities represents a critical gap in current AI healthcare research. Moreover, healthcare coverage in rural regions is shaped not only by infrastructure but also by demographic and clinical context. Providers often depend on high-context, experience-based questioning rather than real-time diagnostics. As such, there is a growing need for AI tools that replicate these diagnostic strategies and function through asynchronous or low-bandwidth communication, including surveys or audio prompts.

Integrating demographic data with technology access indicators, such as overlaying internet coverage maps with population density, can reveal underserved areas and guide the development of localized, low-resource AI systems. In connectivity-limited settings, audio-based consultations and context-aware symptom checkers may offer interim solutions. Ultimately, a shift from disease-centered to people-centered approaches is essential to ensure that AI-driven healthcare solutions are equitable, context-appropriate, and feasible in rural environments.

\subsection{Recommendations for Overcoming Barriers}
To address these challenges and maximize the benefits of AI in rural healthcare, several strategies are recommended:
\begin{enumerate}
\item Model Development and Validation: Prioritize the development of AI models tailored to diverse populations and healthcare settings by integrating diverse datasets, including socio-demographic and omics data. Efforts should focus on reducing algorithmic bias, enhancing diagnostic accuracy, and establishing standardized protocols for clinical evaluation. Future research should particularly address gaps in mental health applications and hybrid human-AI workflows.

\item Technology Integration and Usability: Develop user-friendly interfaces and offline-compatible solutions to address connectivity challenges in rural areas. Efforts should focus on seamlessly integrating AI tools into telemedicine platforms and EHRs to enhance usability and foster collaborative workflows between AI systems and healthcare professionals.

\item Chronic Disease Management: Validate AI interventions for screening, predicting, and managing chronic conditions like diabetes and cardiovascular disease. Research should explore the cost-effectiveness and scalability of these interventions in underserved regions, while expanding to understudied areas such as mental health disorders prevalent in rural communities.

\item Community Engagement and Education: Design and implement community-based education programs to improve health literacy and awareness of AI technologies. Tailored training for healthcare workers is essential to build trust and acceptance of AI solutions, particularly for hybrid models combining AI decision-support with human oversight.

\item Real-World Implementation and Policy Integration: Conduct pilot studies to assess the usability of AI tools and incorporate feedback from diverse stakeholders. Collaborate with policymakers to integrate AI into national health programs, focusing on targeted interventions for underserved populations and longitudinal evaluation frameworks.

\item Longitudinal Studies and Generalizability: Conduct long-term studies to evaluate AI's impact on health outcomes and system resilience. Use longitudinal datasets and causal inference methods to improve the sustainability of AI applications, with particular attention to real-world validation in diverse rural settings.

\item Privacy, Security, and Ethical Considerations: Develop robust frameworks to address data privacy, security, and algorithmic fairness. Transparent policies must guide the ethical deployment of AI in rural settings, including specific provisions for mental health applications and human-AI collaboration models.

\item Population-Specific and Localized Interventions: Tailor AI tools to meet the unique needs of rural populations, emphasizing localized screening strategies and disease-specific interventions. Cultural relevance and community-level solutions are critical for acceptance, particularly for sensitive applications like mental health screening.

\item Cost-Effectiveness and Scalability: Evaluate the economic implications of AI adoption and prioritize scalable, low-cost models to maximize accessibility and sustainability, including hybrid approaches that optimize resource use in low-infrastructure settings.

\item Regulation and Governance: Establish comprehensive regulatory frameworks for the ethical deployment of AI in healthcare. Mechanisms for oversight, transparency, and iterative improvements must be developed to adapt to emerging challenges in human-AI collaboration and mental health applications.

\item Generative AI for Health Literacy and Self-Help: Leverage generative AI tools such as medical-specific large language models (e.g., Med-PaLM, BioGPT) and multilingual chatbots to support self-help, health education, and patient engagement in rural settings. These tools can provide contextually relevant, accessible, and culturally sensitive information, helping to promote awareness, build trust, and increase acceptance of AI-enabled healthcare solutions. Integrating these systems into mobile platforms, while ensuring support for local dialects, can further enhance outreach in areas with low connectivity or limited access to technology, including for mental health support and hybrid care models.
\end{enumerate}

\subsection{Implications of this Review}

The findings of this review underscore AI's transformative potential in addressing rural healthcare challenges. By enhancing diagnostic precision, optimizing resource allocation, and improving patient outcomes, AI can help overcome infrastructural and workforce deficiencies. Tailored AI solutions that consider rural populations' unique epidemiological, geographic, and socioeconomic characteristics are critical. Effective implementation requires collaboration among clinicians, technologists, policymakers, and community representatives. Healthcare professionals must receive targeted training and continuing education to adopt AI tools effectively. These programs should address knowledge gaps, focus on practical applications, and build confidence in AI as a decision-support tool.

Policymakers and healthcare leaders must prioritize scalable, low-cost AI solutions while establishing regulatory frameworks to ensure ethical and transparent use. Future research should refine AI models to address bias, interpretability, and scalability and explore their integration with complementary technologies such as mobile health platforms and wearable devices. By systematically addressing these priorities, stakeholders can ensure AI technologies' ethical, inclusive, and efficient deployment. This proactive approach has the potential to bridge healthcare disparities, enhance equity, and foster a resilient, patient-centred healthcare ecosystem in underserved rural regions.


\section{Conclusion} \label{con}
Integrating AI in rural healthcare delivery holds transformative potential to address longstanding challenges, including workforce shortages, limited infrastructure, and inequitable access to essential services. Innovative diagnostics, chronic disease management, and public health applications have demonstrated AI's capacity to improve patient outcomes and optimize resource allocation in underserved areas. However, significant barriers, such as data limitations, infrastructural constraints, and ethical considerations, impede widespread adoption. Tailoring AI solutions to the unique needs of rural populations and prioritizing inclusivity, transparency, and community engagement is essential for effective implementation. Addressing infrastructural barriers, fostering interdisciplinary collaboration, and establishing ethical frameworks remain critical steps toward equitable AI deployment. Robust policy support and localized interventions are vital to bridging the gap between technological advancements and real-world healthcare delivery in resource-constrained settings. Future efforts should focus on scaling these innovations sustainably, ensuring accessibility, and aligning them with the specific needs of rural communities. By proactively addressing these challenges, AI can drive a paradigm shift in rural healthcare delivery, reducing disparities and strengthening healthcare systems.

\section*{Data Availability Statement}
The datasets used and/or analysed during the current study are available from the corresponding author on reasonable request
\section*{Funding Declarations}
No funding was received for this project.
\section*{Author Contributions Statement}
K.B., D.V., and H.E.H. conceived the initial idea. D.V. and Z.L. conducted the literature search, study selection, and data extraction. K.B. and D.V. synthesized the results and drafted the manuscript. H.E.H., K.R., and H.K. contributed to interpreting the findings. S.R. and P.M.S. reviewed and edited the manuscript. All authors revised and approved the final manuscript.
\section*{Competing Interests Statement}
The authors declare no competing interests.
\section*{Ethics, Consent to Participate, and Consent to Publish declarations}
Not applicable, as this study is a review of publicly available literature and does not involve human participants, identifiable data, or the collection of new data requiring ethical approval or consent.

\end{document}